\documentclass[twocolumn,iop,numberedappendix,appendixfloats]{openjournal}

\usepackage[table, dvipsnames,svgnames,x11names]{xcolor}
\usepackage[colorlinks=true, linktocpage, linkcolor={blue!60!black}, citecolor={blue!60!black}, urlcolor={blue!60!black}]{hyperref}
\usepackage{scrextend}
\usepackage{amsmath}
\usepackage{amssymb}
\usepackage{graphicx}
 \usepackage{booktabs}

\usepackage{txfonts}
\usepackage{enumitem}

\def\ESO{1}
\def\LEIDEN{2}
\def\BrookhavenLab{3}
\def\BerkleyLab{4}
\def\UOslo{5}
\def\UPENN{6}
\def\AZUR{7}
\def\UTRIEST{8}
\def\INAFTRIEST{9}
\def\IFPU{10}
\def\UofMi{11}
\def\CFA{12}
\def\TUM{13}
\def\ORIGINS{14}

\begin{document} 
    \journalinfo{The Open Journal of Astrophysics}
    \submitted{submitted; Nov 25, 2024, accepted; \today}

    \title{\texttt{maria}: A novel simulator for forecasting (sub-)mm observations}

    \author{ J.~van Marrewijk\altaffilmark{\ESO, \LEIDEN},
            T.~W.~Morris\altaffilmark{\BrookhavenLab, \BerkleyLab},
            T.~Mroczkowski\altaffilmark{\ESO},
            C.~Cicone\altaffilmark{\UOslo},
            S.~Dicker\altaffilmark{\UPENN},
            L.~Di Mascolo\altaffilmark{\AZUR,\UTRIEST,\INAFTRIEST,\IFPU},
            S.~K. Haridas\altaffilmark{\UPENN},
            J.~Orlowski-Scherer\altaffilmark{\UPENN},
            E.~Rasia\altaffilmark{\INAFTRIEST, \IFPU, \UofMi},
            C.~Romero\altaffilmark{\CFA},
            and 
            J. W\"urzinger\altaffilmark{\TUM, \ORIGINS}
    }
    
    \affiliation{\altaffilmark{\ESO} European Southern Observatory (ESO), Karl-Schwarzschild-Strasse 2, Garching 85748, Germany}
    \affiliation{\altaffilmark{\LEIDEN} Leiden Observatory, Leiden University, P.O. Box 9513, 2300 RA Leiden, The Netherlands}
    \affiliation{\altaffilmark{\BrookhavenLab} Brookhaven National Laboratory, Upton, NY 11973, USA}
    \affiliation{\altaffilmark{\BerkleyLab} Lawrence Berkeley National Laboratory, Berkeley, CA 94720, USA}
    \affiliation{\altaffilmark{\UOslo} Institute of Theoretical Astrophysics, University of Oslo, P.O. Box 1029, Blindern, 0315 Oslo, Norway}
    \affiliation{\altaffilmark{\UPENN} University of Pennsylvania, 209 S. 33rd St., Philadelphia, PA 19014, USA}
    \affiliation{\altaffilmark{\AZUR} Laboratoire Lagrange, Université Côte d’Azur, Observatoire de la Côte d’Azur, CNRS, Blvd de l’Observatoire, CS 34229, 06304 Nice cedex 4, France}
    \affiliation{\altaffilmark{\UTRIEST} Astronomy Unit, Department of Physics, University of Trieste, via Tiepolo 11, Trieste 34131, Italy}
    \affiliation{\altaffilmark{\INAFTRIEST} INAF - Osservatorio Astronomico di Trieste, via Tiepolo 11, Trieste 34131, Italy}
    \affiliation{\altaffilmark{\IFPU} IFPU - Institute for Fundamental Physics of the Universe, Via Beirut 2, 34014 Trieste, Italy}
    \affiliation{\altaffilmark{\UofMi} Department of Physics; University of Michigan, 450 Church St, Ann Arbor, MI 48109, USA}    
    \affiliation{\altaffilmark{\CFA} Center for Astrophysics, Harvard \& Smithsonian, 60 Garden Street, Cambridge, MA 02138, USA}
    \affiliation{\altaffilmark{\TUM} Technical University of Munich, Arcisstrasse 21, 80333 Munich, Germany}
    \affiliation{\altaffilmark{\ORIGINS} Excellence Cluster ORIGINS, Boltzmannstrasse 2, 85748 Garching, Germany}

    \begin{abstract}
       Submillimeter/millimeter-wave single-dish telescopes offer two key advantages compared to interferometers: they can efficiently map larger portions of the sky, and they can recover larger spatial scales. Nonetheless, fluctuations in the atmosphere, the dominant noise source in ground-based observations, limit the accurate retrieval of signals from astronomical sources.
       We introduce a versatile, user-friendly simulator to optimize scanning strategies and instrument designs to efficiently reduce atmospheric noise and filtering effects. We further use this tool to produce synthetic time streams and maps from hydrodynamical simulations, enabling a fair comparison between theory and reality.
       To generate synthetic time-ordered data, we developed a multi-purpose telescope simulator called \texttt{maria}, which implements a suite of telescope and instrument designs intended to mimic current and future facilities. Each mock observatory scans through the atmosphere in a configurable pattern over the celestial object. We generate evolving and location-and-time-specific weather for each of the fiducial sites using a combination of satellite and ground-based measurements. While \texttt{maria} is a generic virtual telescope, this study specifically focuses on mimicking broadband bolometers observing at $\approx100~$GHz.
       To validate our virtual telescope, we compare the mock time streams with real MUSTANG-2 observations and find that they are quantitatively similar by conducting a $k$-sample Anderson-Darling test resulting in a $p-$value of $p<0.001$. 
       Subsequently, we image the time-ordered data to create noise maps and realistic mock observations of clusters of galaxies for both MUSTANG-2 and an instrument concept for the 50~m Atacama Large Aperture Submillimeter Telescope (AtLAST).
       Furthermore, using \texttt{maria}, we find that a 50~m dish provides the highest levels of correlation of atmospheric signals across adjacent detectors compared to smaller apertures (e.g., 42-cm and 6-m survey experiments), facilitating removal of atmospheric signal on large scales.  
       We demonstrated the usage of \texttt{maria} for facilitating detailed forecasts for frontier instruments, maximizing their scientific output.
    \end{abstract}
    

    \shorttitle{Simulating (sub-)mm observations}
    \shortauthors{van Marrewijk, Morris, et al.}  
    
    \maketitle

\section{Introduction}

    The most sensitive (sub-)millimeter [hereafter, (sub-)mm] user facility yet built is the Atacama Large Millimeter/submillimeter Array (ALMA), which outperforms current single-dish telescopes and other interferometers in terms of point source sensitivity due to its large collecting area and excellent site. However, interferometers like ALMA are optimized to achieve sub-arcsecond resolution in the (sub-)mm, but this comes at the expense of being unable to constrain large spatial scales, which are probed by short antenna spacings. This is simply due to the minimum distance between two antennas needed to avoid collisions while tracking a source. This limit, known as the shadowing limit, prevents zero-spacing information and results in a biased recovery of the total integrated flux. Previous works have shown that even at scales the size of the synthesized beam (i.e., the interferometric resolution element), only a fraction of the true flux is recovered \citep[][]{Plunkett2023, Bonanomi2024}. Simulations indicate that to address the loss of information at larger angular scales and to provide good overlap in the Fourier domain, one should rely on single-dish telescopes with a dish size of at least three times the size of the interferometric array element. This is necessary to provide the complete Fourier sampling required for high spatial dynamic range and unbiased image reconstruction \citep{Frayer2017, Plunkett2023}. 

    On the other hand, large single-dish telescopes have their own observational limitations. For example, due to the faintness of the signal compared to the Earth's atmospheric emission, drifts in detector gain and response, and the fact that most bolometer-based instruments are not set up to perform an absolute temperature calibration, bolometric arrays require differential measurements (e.g., going on and off source) to obtain an unbiased flux estimate. Single-dish user-facility telescopes like the Green Bank Telescope (GBT) therefore adopt a Lissajous daisy scanning pattern \citep[e.g.][]{Romero2020}. These scanning trajectories impact the observed sky and introduce filtering effects, such as a higher sensitivity in the center compared to the outskirts of the observation.

    Removing atmospheric contamination from differential measurements poses a significant challenge for ground-based telescopes. Simpler methods involve subtracting a common mode from the data or high-pass filtering the time-ordered data (TODs, also referred to as time streams) to exclude fluctuations below a certain threshold frequency. More sophisticated techniques may take into account non-trivial correlations between different detectors, how the fluctuations change over time, and how TODs cohere across time due to the periodicity of the scan pattern. \citet{morris2022} has shown that the atmosphere exhibits these non-trivial correlations, which determine how effective any given mapmaking strategy will be. However, since the data comprise a superposition of astronomical signals, atmospheric influences, and detector noise, a faithful observation simulator can help inform us how different filtering and data reduction techniques impact mapmaking and the recovery of celestial signals.
    
\begin{figure}[t]
        \centering
        \includegraphics[width =\hsize]{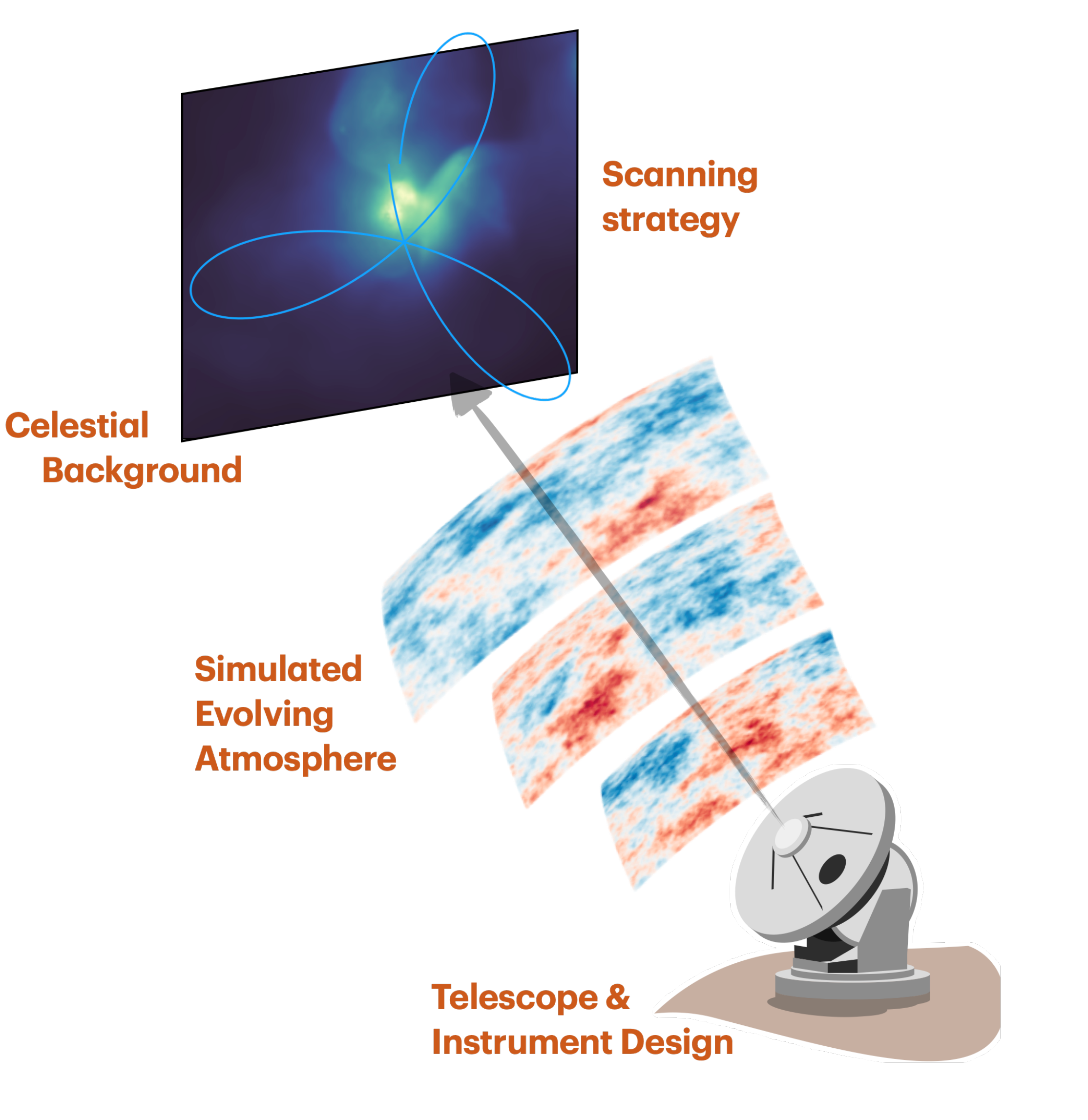}
        \caption{Sketch on how \texttt{maria} operates. The virtual telescope consists of four parts: A celestial background, a location and time-specific evolving atmosphere, a telescope and instrument design, and a scanning strategy.}
        \label{fig:overview}
    \end{figure}

    In this work, we adapt {\tt maria}\footnote{\label{footnote:maria}See \url{https://thomaswmorris.com/maria/}} to perform as a virtual single-dish observational simulator, designed to closely mimic real-world conditions. We leveraged this tool to generate realistic synthetic TODs and maps from simulations, enabling a one-to-one comparison between simulated objects evolved in a cosmological environment, offering a more complex and accurate description than simple analytic models, and observational data. The tool is constructed as follows: we employ a predefined telescope and detector array design to scan in a daisy or back-and-forth azimuthal scan pattern through the ever-changing atmosphere over a celestial body. We visualized these components in a sketch shown in Figure~\ref{fig:overview}. This approach allows studies on optimizing telescope design and scanning strategies. Thus, \texttt{maria} functions similarly to codes such as \texttt{X-MAS} \citep{Gardini2004, Rasia2008}, \texttt{Phox} \citep{Biffi2011}, \texttt{pyXSIM} \citep{ZuHone2016}, and \texttt{SOXS} \citep{ZuHone2023} written to mimic X-ray observations, \textsc{casa} \texttt{simobserve} \citep{CASA2022} regarding interferometric observations, and TOAST \citep{Puglisi2021} which is primarily used for Cosmic Microwave Background (CMB) survey forecasts.
    
    The current implementation of \texttt{maria} is flexible and can be customized to various types of instruments, such as heterodyne and integrated field units, to generate mock datacubes with spectral information. However, in this study specifically, we limit \texttt{maria} to generate synthetic broadband bolometric observations. In particular, we restrict \texttt{maria} three-fold: (i) We mimic only two facilities, of which one already exists (the GBT with the MUSTANG-2 instrument) which allows testing the {\tt maria} output with real observations, and a future facility, namely the 50~meter Atacama Large Aperture Submillimeter Telescope (AtLAST; \citealt{Ramasawmy2022, mroczkowski2023, Booth2024, Mroczkowski2024, Reichert2024})\footnote{See also: \url{https://atlast-telescope.org}} with a possible bolometer instrument concept.\footnote{We note that AtLAST will host a multitude of different instruments, also including spectroscopic receivers.} The tool itself includes configuration files to easily simulate more facilities, such as the Large Millimeter Telescope (\citealp[LMT;][]{Kaercher2000}), as well as current CMB survey experiments such as the South Pole Telescope (\citealp[SPT;][]{Carlstrom2011, Bleem2015, Bleem2020}) and the Atacama Cosmology Telescope \citep[ACT;][]{Swetz_2011, Thornton_2016, henderson2016advanced}, imminent ones like Simons Observatory \citep[SO,][]{Ade2019} and CCAT-prime (Fred Young Sub-mm Telescope or FYST; \citealt{CCAT2023}), and future ones like CMB-S4 \citep{CMBS42016}. (ii) We focus on the $\nu \sim 100$~GHz regime, which is the high-frequency regime for the GBT, and the low-frequency regime for a telescope such as AtLAST. We note that {\tt maria} can also simulate the atmosphere at higher frequencies up to where the atmosphere is still transmissive well into the sub-mm. (iii) Finally, even though \texttt{maria} is generic and can be employed for a wide variety of science cases, we focus here on a specific one: the intracluster medium seen via the thermal Sunyaev-Zeldovich (SZ; \citealt{Sunyaev1970, Sunyaev1972}) effect. Briefly, the thermal SZ effect is a spectral distortion to the CMB, presents as a faint, extended decrement at frequencies $\nu \lesssim 218$~GHz, and is a probe of gas thermal energy (see, e.g., \citealt{Mroczkowski2019} for a review, as well as \citealt{DiMascolo2024} for a more comprehensive overview of key AtLAST SZ science cases).  

    The remainder of this work is structured as follows: in Section~\ref{sec:method} we provide a detailed description of the simulator's construction, elucidating its four constituent components. Section~\ref{sec:valid} demonstrates the simulator's ability to replicate existing MUSTANG-2 observations and offers an example of how it can be employed to assess the feasibility of scientific objectives. Moving forward, Section~\ref{sec:AtLAST} delves into the specifics of forecasting for AtLAST. Lastly, Section~\ref{sec:conclusion} offers a comprehensive summary and discussion of our study.
    
\section{A generic virtual single dish telescope}\label{sec:method}

    The simulator we built is a continuation of the atmospheric modeling work done by \citet{morris2022, Morris2024}. Therefore, we refer to those works and references therein for an in-depth discussion of atmospheric modeling. Here, we briefly describe how the \texttt{maria} simulation works. The simulator consists of five main components: 1) a telescope and instrument design, 2) a scanning strategy, 3) an input astronomical map, 4) a site and time location to generate atmospheric weather conditions, and 5) additional additive noise and detector inefficiencies. These components are visualized in Figure~\ref{fig:overview} and will be discussed separately in the following five subsections. For the full overview of all the parameters and corresponding tutorials, we refer to the \texttt{maria} documentation.\footref{footnote:maria}
    With this setup, we can directly link fundamental design choices, such as primary size and the number of detectors, to a scientific output. Thus, we established a tool for communication between instrument scientists and observers, which is crucial for the development of future facilities.

    \subsection{Instrument configurations \& Telescope designs}

        In \texttt{maria}, each simulation is characterized by a set of fundamental parameters regarding both the telescope and instrument design. Specifically, we use the primary size, the field of view (FoV), the number of detectors, and the azimuthal and elevation bounds of the telescope for modeling the simulations. Further, \texttt{maria} positions detectors in the array evenly to fill the FoV in a hexagonally packed configuration. In the current implementation, each detector is specified with a default frequency band with a flat response function (e.g., a flat passband from 140-160~GHz, centered at 150~GHz). \texttt{maria} also hosts the ability to import passbands.

        In this study, we mimic two telescope and instrument designs: the operational MUSTANG-2 instrument (which uses bolometers) mounted on the GBT and a possible small-scale broadband continuum instrument concept for AtLAST. As noted in Section \ref{sec:AtLAST}, the AtLAST concept presented here is limited by our own computational considerations rather than expected observatory capabilities. Table~\ref{tab:design_table} provides an overview of the different designs of the two telescopes and instruments discussed in the next two subsections. It also includes parameters, which we will introduce in the remainder of this section.

        \subsubsection{MUSTANG-2 on the GBT}

            MUSTANG-2 is a 223-detector bolometer array on the 100~meter Green Bank Telescope (GBT), located in Green Bank, West Virginia at an elevation of 818\,m above sea level.
            MUSTANG-2 achieves a $8\farcs5$ full width at half maximum (FWHM) resolution with an instantaneous FoV of $4\farcm2$ centered around 93~GHz, with a continuum bandwidth of $\approx$30~GHz. The detectors, which are horn-coupled, are spaced at $1.9~f-\lambda$. For further information, see \cite{Dicker2014}. The placement of MUSTANG-2 detectors on the sky is depicted in the left panel of Figure~\ref{fig:setup}. 
            \texttt{maria} includes a set of configuration files with hard-coded parameters to match MUSTANG-2 as well as other existing and upcoming telescope designs. However, as documented, any of these parameters can be reconfigured.

        \subsubsection{AtLAST}\label{sec:design_atlast}

            AtLAST \citep{Klaassen2020, Ramasawmy2022} is a concept for a next-generation high-throughput sub-mm telescope to be sited high in the Atacama Desert of Chile, on Llano de Chajnantor at an elevation of approximately 5100 m above sea level (i.e., the same plateau as that on which ALMA resides). AtLAST underwent a design study, funded by the European Commission,\footnote{See \url{https://cordis.europa.eu/project/id/951815}.} and has been selected for a new infrastructure development grant, also funded by the European Commission, starting in January 2025. Among other activities, AtLAST will continue to improve its technology readiness levels and perform scientific forecasts using, for example, the {\tt maria} code presented here.  
            With a surface accuracy of 20~$\mu$m, AtLAST will be capable of observations up to 950 GHz (i.e., \ in the 350~$\mu$m atmospheric window accessible in top octile conditions there).
            AtLAST will include dedicated instrument spaces for six extremely large receivers, two of which can completely access the full 4.7-meter focal plane corresponding to the 2$^\circ$ FoV \citep[see][]{Gallardo2024, Kiselev2024, Puddu2024, Reichert2024, Mroczkowski2024}. 

            \begin{table}[t]
                \centering
                \begin{tabular}{@{}lll@{}}
                    \toprule
                        Telescope Design & GBT & AtLAST  \\
                    \midrule
                        Primary size              & 100 m       & 50 m \\
                        Resolution at 90~GHz      & $8\farcs5$ & $16\arcsec$ \\
                        Field of View$^\dagger$   & $15\arcmin$ & $2^\circ$ \\  
                        Site location             & Green Bank  & Atacama Desert \\
                        ~ & WV, USA & Chile\\
                        Elevation                 & 818 m       & 5100 m \\
                        Surface accuracy          & 230 $\mu$m  & 20 $\mu$m \\
                        Frequency coverage        & 1-116 GHz            & $\approx 30 - 950$~GHz\\
                        Scanning speed    & $\approx 50\arcsec/s$ & $\approx 3^\circ/s$ \\
                    \bottomrule \\
                        Simulated instrument$^\star$ & & \\
                    \midrule
                        Frequency coverage      & 78-108~GHz & 66-118~GHz\\
                        Field of View           & $4\farcm2$ & $0.25^\circ$ \\ 
                        scanning radius    & $4\arcmin$ & $ 0.25^\circ$ \\
                        Read-out rate           & 100~Hz & 225~Hz\\
                        Detector Count          & 223    & 3000 \\ 
                        Optical efficiency      & 0.3 & 0.3 \\
                        Illumination efficiency & 0.8 & 0.8 \\
                        White noise level       & 1.7 mK\,$\sqrt{\rm s}$ & 0.266 mK\,$\sqrt{\rm s}$\\
                        PWV RMS                 & 0.5-5\% & 1-10\% \\
                    \bottomrule
                \end{tabular}
                \smallskip
                \caption{\centering Instrument and telescope design differences between the Green Bank Telescope (GBT) and the 50~meter Atacama Large Aperture Submillimeter Telescope (AtLAST). $^\star$ Here, we simulate MUSTANG-2 on the GBT and a small-scale plausible continuum receiver for AtLAST. For a more in-depth description of the used parameters for the latter design, we refer to Section~\ref{sec:AtLAST}. $^\dagger$ Here, we estimate the full field of view of the GBT. The detector array FoV of MUSTANG-2 is limited by the cryostat size, which in turn is limited by its location in the GBT instrument turret.  The full $2^\circ$ FoV of AtLAST is accessible for two of the 6 planned instrument locations \citep[see][]{mroczkowski2023, Mroczkowski2024}.}
                \label{tab:design_table}
                \vspace{-1mm}
            \end{table}
            
    \subsection{Scanning strategies}
    
    \begin{figure*}[t]
        \centering
        \includegraphics[trim={0 0 0 0cm}, clip, width = 0.9\textwidth]{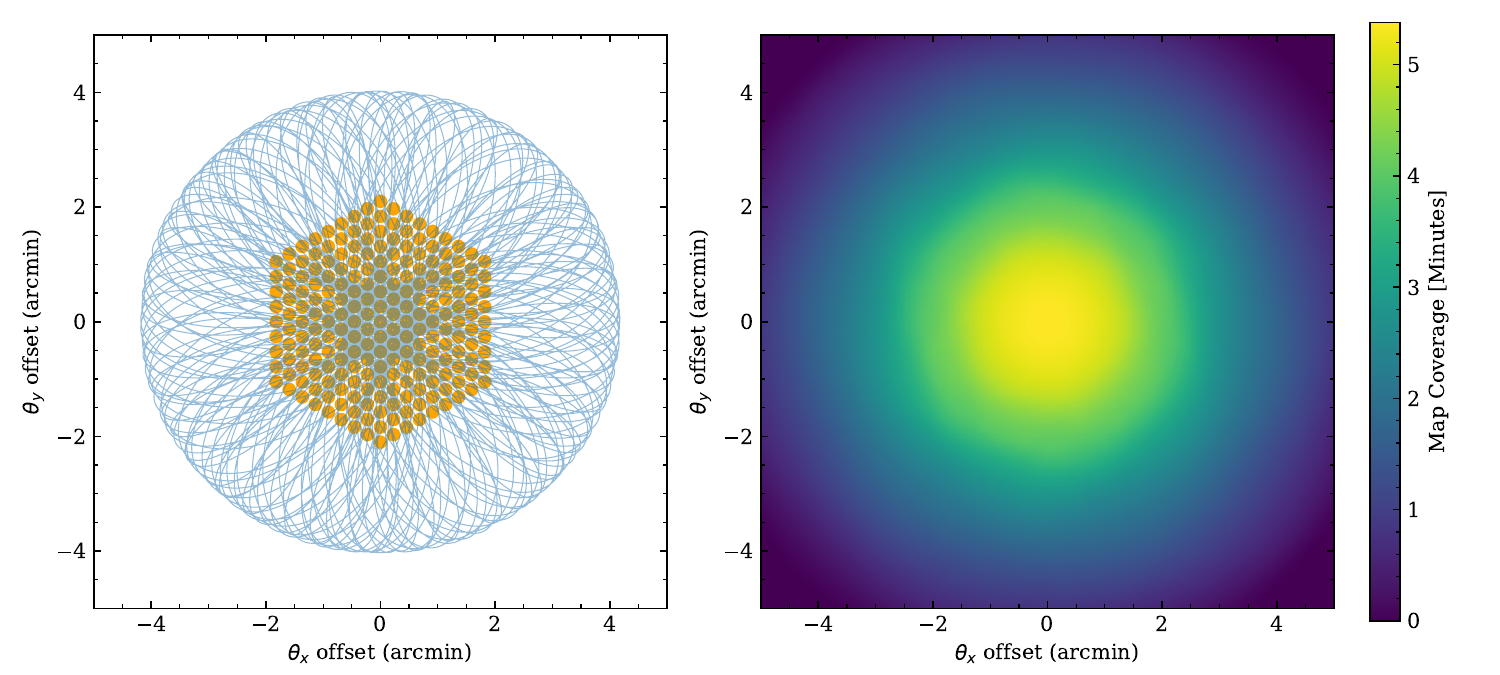}
        \caption{Example of a realistic scanning strategy used to mimic a MUSTANG-2 observation with \texttt{maria}. Here, we scan for 45 minutes with a read-out rate of 50~Hz and a maximum scan velocity of $51\arcsec \rm s^{-1}$, which occurs at the center of the Lissajous daisy scan. The left panel shows the instantaneous FoV of all detectors in the 93 GHz band. This array will be moved over the sky in the daisy pattern illustrated in blue. This scanning pattern leads to a higher sensitivity in the center than in the outskirts, as illustrated on the right panel, which shows how long each resolution element (a $8\farcs5$ beam) is covered by a detector. This hit map is smoothed with the diffraction-limited beam of the telescope. The pixel size of the map is set to $2\arcsec$, the value to Nyquist sample the MUSTANG-2 beam.}
        \label{fig:setup}
    \end{figure*}

        After specifying the instrument configuration and telescope design in \texttt{maria}, we define a scanning strategy that determines the pointing of the virtual telescope throughout the observation. In this work, we use a double Lissajous daisy (or ``daisy scan'').
        The double Lissajous daisy is defined as:

        \begin{equation}
            \begin{aligned}
                \Vec{x_{\rm outer}} &= r_{\rm outer}\sin{\left(\phi_{\rm outer}\right)}~\exp{\left(i~\phi_{\rm outer}/{\rm n_{\rm petals}}\right)}, \\
                \Vec{x_{\rm inner}} &= r_{\rm inner }\sin{\left(\phi_{\rm outer}+\pi/2\right)}~\exp{\left(i~\phi_{\rm inner}/{\rm n_{\rm petals}}\right)}, \\
                \Vec{x_{\rm scan}} &= \Vec{x_{\rm outer}} + \Vec{x_{\rm inner}}, \\
                \rm RA,~Dec &= 	\mathbb{R}\left[\Vec{x_{\rm scan}}\right],~\mathbb{I}\left[\Vec{x_{\rm scan}}\right],
            \end{aligned}
        \end{equation}

        \noindent with the $r_{\rm outer}$ the scanning radius of the daisy scan, $\phi_{\rm outer}$ the phase which is defined as $\phi_{\rm outer} = t~v_{\rm scan}/r_{\rm outer}$, where the $t$ is the time stamp and $v_{\rm scan}$ the scanning velocity. 
        We set $r_{\rm inner} = 0.15 r_{\rm outer}$ and $\phi_{\rm innter} = \sqrt{2} \phi_{\rm outer}$, and $\rm n_{\rm petals} = 10/\pi$. We find this set of parameters to produce a daisy pattern which is ideal for MUSTANG-2-like observations. The resulting daisy scan is shown on the left panel of Figure~\ref{fig:setup}. Using this scanning method, the detector array traverses the sky in a daisy pattern, passing close to the center after every pedal. This approach results in more frequent coverage in the middle of the field compared to the outskirts, leading to higher central sensitivity, as shown on the right side of Figure~\ref{fig:setup}. That panel shows how long each pixel is covered by a detector during 45 minutes of scanning.

        For the daisy scanning method, we specify the maximum scanning velocity (defined as the velocity at the center of the daisy scan), read-out rate,  scanning radius, offset parameter, and pointing center. However, other scanning strategies, such as a simple `stare' or a back-and-forth azimuthal scan (e.g., a constant elevation scan), are also implemented in \texttt{maria}. Different scanning strategies lead to different types of sensitivity maps. For instance, when targeting clusters of galaxies, which are known as large extended objects, the GBT/MUSTANG-2 changes its scanning strategy by increasing the scanning radius of the daisy scan and changing the pointing in a box-like pattern after each scan by an arcminute to spread the sensitivity of the center to a wider area. These differences in scanning strategies can easily translate to a central sensitivity difference of a factor of $\sim 1.5$. Thus, one is free to choose how to optimize the observation strategies for a given science goal and can do so through \texttt{maria}. 

    \subsection{Input astronomical map}

        The third component is the astronomical data from which we create a synthesized observation. We employ an approach similar to that of \textsc{casa} \texttt{simobserve}, a tool used broadly for simulating interferometric observations. As an input file, \texttt{maria} requires a flexible image transport system ({FITS}; \citealt{Wells1981}) file in units of Jy/pixel or in Rayleigh-Jeans Kelvin ($\rm K_{RJ}$). Also, the pixel size must be specified in units of degrees or radians. The input file will be interpreted as the true background signal.
        
        Often in the millimeter-wave regime, astronomical signals are contaminated by fore- and background sources, like the CMB, the Galaxy, and the Cosmic Infrared Background (CIB), where the latter is generally the unresolved confusion-limited sub-mm/FIR emission dominated by dusty galaxies from different epochs, peaking between $z\sim 0.5- 5$ \citep[see, i.e.,][]{Devlin2009, MacCrann2024, Madhavacheril2024}. In future implementations of \texttt{maria}, there will be an option to include these components through a power spectra modelization, which will be jointly mapped with the atmosphere. However, for the current version, we recommend manually adding these contamination sources to the input file when making forecasts for survey experiments, such as detecting the primary CMB or when performing line intensity mapping experiments.

    \subsection{Atmospheric modeling}

        The atmospheric modeling in \texttt{maria} is the backbone of the corruption model of the simulation. Therefore, this subsection details how the atmospheric modeling is set up and implemented. In particular, we discuss which weather data is used to generate local-and-time specific initial conditions (Section~\ref{sec:weather_data}), how we generate atmospheric emission (Section~\ref{sec:atmos_emission}) and corresponding turbulence from the initial conditions (Section~\ref{sec:turbulence}), how we let the atmosphere evolve with time while scanning over it (Section~\ref{sec:evolving}), and how all this is implemented in \texttt{maria} (Section~\ref{sec:atmos_imp}). For a more comprehensive overview of how the atmospheric model is built and how it compares with ACT observations, we refer to works of \citet{morris2022, Morris2024}.

        \subsubsection{Weather data}\label{sec:weather_data}

            To simulate weather conditions at astronomical sites, \texttt{maria} uses reanalysis data from the European Centre for Medium-Range Weather Forecasts \citep[ECMWF,][where the fifth-generation atmospheric reanalysis is known as ERA5]{ERA5}.\footnote{See \url{https://thomaswmorris.com/maria/sites} for a list of supported sites.} 
            Reanalysis data is a mixture of weather observations with past short-range forecasts rerun with modern weather forecasting models from \citet{ERA5}. It further has the advantage of providing a characterization of the entire atmospheric column, which can often be quite different from ground-based weather measurements. 

        \subsubsection{Atmospheric emission and opacity}\label{sec:atmos_emission}

            \begin{figure}[t!]
                \centering
                \includegraphics[width = 0.5\textwidth]{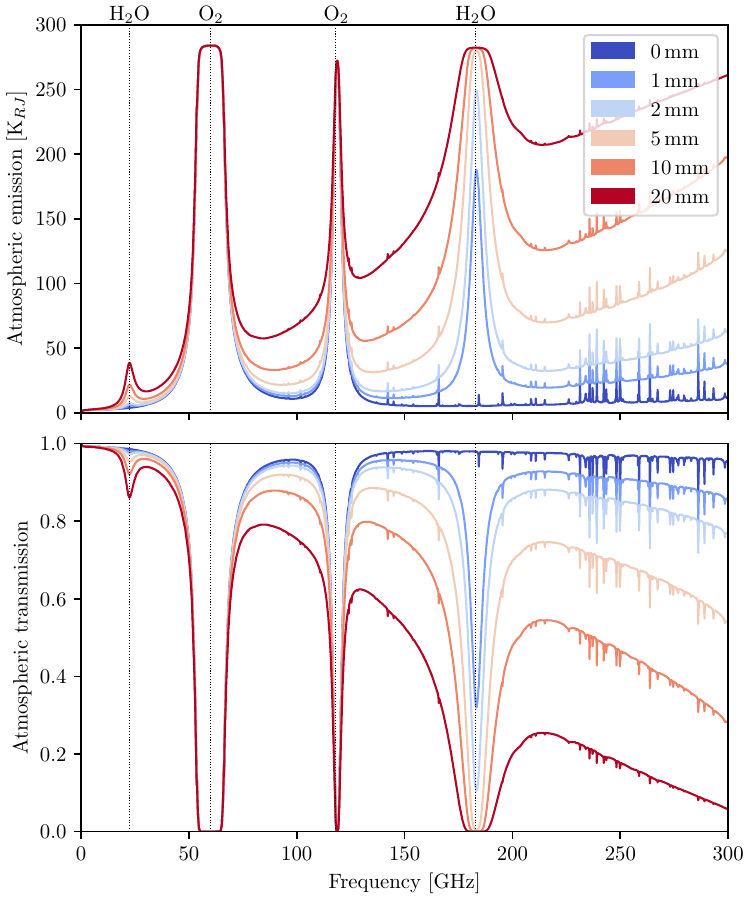}
                \caption{Emission and transmission spectra of the atmosphere at the Green Bank site for different levels of precipitable water vapor (PWV) computed using the \texttt{am} code \citep{am:2018}. The four most significant spectral lines, due to water and molecular oxygen, are identified on the top axis. On time scales longer than a few seconds, the fluctuations in PWV due to spatial fluctuations in temperature and humidity generally dominate the noise temperature fluctuations observed (see Figs.~\ref{fig:weather}~\&~\ref{fig:shape}.)}
                \label{fig:atm_spectrum}
            \end{figure}

            As the dominant source of fluctuation in atmospheric emission and opacity for submillimeter telescopes is precipitable water vapor (PWV), we simulate atmospheric fluctuations by perturbing the line-of-sight water vapor of the receiver. 
            We do this by generating a set of layers of turbulent fluctuations and scaling them appropriately according to site-specific weather data derived from ERA5.
            Typical minute-to-minute variations in PWV are on the order of a few percent of the total \citep{Morris2024}, though the relative fluctuation is a variable parameter in \texttt{maria}. We compute the atmospheric emission and opacity spectra using the vertical profiles of temperature, humidity, ozone, and wind velocity using the \texttt{am} atmospheric model \citep{am:2018} for a given site, altitude, and observing elevation. Samples of these spectra for different levels of PWV at the Green Bank site are shown in Figure~\ref{fig:atm_spectrum}.

        \subsubsection{Modeling turbulence}\label{sec:turbulence}

            Atmospheric emission comprises a large, constant brightness from the atmosphere (which is almost always subtracted out of the timestream) but also a smaller time-dependent brightness due mostly to fluctuations in atmosphere temperature and water vapor density. Due to turbulent mixing between atmospheric layers of different temperature and humidity, local turbulent variations introduce spatial and temporal fluctuations in the atmosphere; see \citealt{tatarski} for a statistical description of this phenomenon. For small angular separations, \citet{morris2022} showed that the angular covariance of turbulent fluctuations in the line-of-sight atmospheric emission $\delta T(\theta)$ can be modeled as:
    
            \begin{equation}\label{eq:angular_turbulence}
                \Big \langle \delta T(\theta_0) \delta T(\theta_0 + \theta) \Big \rangle \propto \int_0^\infty \sigma^2(z) \Bigg ( \frac{z \theta}{r_0} \Bigg )^{5/6} K_{5/6} \Bigg ( \frac{z \theta}{r_0} \Bigg )  dz,
            \end{equation}
    
            \noindent where $\theta$ is the angular separation on the sky, $z$ is the distance from the observer, $K_{\nu}(\cdot)$ is the modified Bessel function of the second kind of order $\nu$, and $\sigma^2(z)$ is the scaling of atmospheric brightness fluctuations as a function of distance from the observer, and $r_0$ is the turbulent outer scale.\footnote{An analogous relation holds for opacity fluctuations.} This equation can be interpreted as a series of ``integrated'' layers of turbulence with a modified spectrum. For small angular separations, we recover a $5/3$ power law in the structure function which arises from:

            \begin{equation}
                 \lim_{\theta \to 0} \big [ C(0) - C(\theta) \big ] \propto \theta^{5/3},
            \end{equation}
            
            \noindent where $C(\theta) = \langle \delta T(\theta_0) \delta T(\theta_0 + \theta) \rangle$ is the turbulent covariance function in Eq.~\eqref{eq:angular_turbulence}. This $5/3$ power law has been observed in, among others, \cite{consortini1972choice}, \cite{conan2000analytical}, \cite{tokovinin2002differential}, and \cite{morris2022}. We use the value of $500$\, meters for the turbulent outer scale $r_0$, which comes from measurements done by \citet{errard2015} and \citet{Morris2024}. We further assume a ``frozen-in" model for the turbulence \citep{lay:1997}, meaning that fluctuations are only induced by the motion along the line-of-sight with respect to the atmosphere as translated by the wind. For each detector $i$, we can compute its atmosphere-relative coordinates as 
    
            \begin{equation}\label{eq:angular_coords}
                \theta_i(t) = \Delta \theta_i + \theta_\text{scan}(t) + \int_0^t \omega(t) dt,
            \end{equation}
    
            \noindent where the first term is the detector offset from the boresight, the second is the motion of the boresight from the scan, and the third term represents the cumulative motion of the atmosphere. The angular wind velocities are derived from reanalysis profiles, which \cite{morris2022} shows agree with what the telescope sees. Following this model, \texttt{maria} simulates layers of turbulence: for each depth of 10 layers between a height of $z = 500$ and $z = 5000$ meters away from the receiver, \texttt{maria} computes the angles that the beams will pass through and simulates turbulence to cover it.

            \subsubsection{Modeling the evolving atmosphere}\label{sec:evolving}

                The \texttt{maria} code simulates turbulence using an auto-regressive version of the covariance matrix method (see e.g., \citealt{assemat2006method}). This method is preferred to Fourier-based methods (see e.g., \citealt{jia2015simulation}) for the simulations in this work, as they employ scans over a small area on the sky with long integration times. This means that the set of angular positions as described by Eq.~\ref{eq:angular_coords} will comprise a long, thin strip of atmosphere which is much more efficiently generated with the former method. The covariance matrix method assumes that perturbations in atmosphere temperature and PWV due to turbulent mixing are described by a stationary Gaussian process with some covariance matrix. Generating a realization of such a process requires inverting the covariance matrix, which scales with cubic complexity in the number of points $n$ and rapidly becomes computationally infeasible for $n \gtrsim 10^4$. We can instead generate a realization iteratively by conditioning a posterior on a sparse sampling of already-realized points and sampling from it. Considerations of sampling methods for balancing performance and realism are addressed in \citet{Morris2024}, along with the trade-offs of different methods of turbulent generation. We compute the angular covariance as described by Eq.~\ref{eq:angular_turbulence}, and adjust the generated turbulent distribution to recover the mean and variance matching the simulation parameters. 
        
            \subsubsection{Implementation in \texttt{maria}}\label{sec:atmos_imp}

                \begin{figure*}[t!]
                    \centering
                    \includegraphics[width = \textwidth]{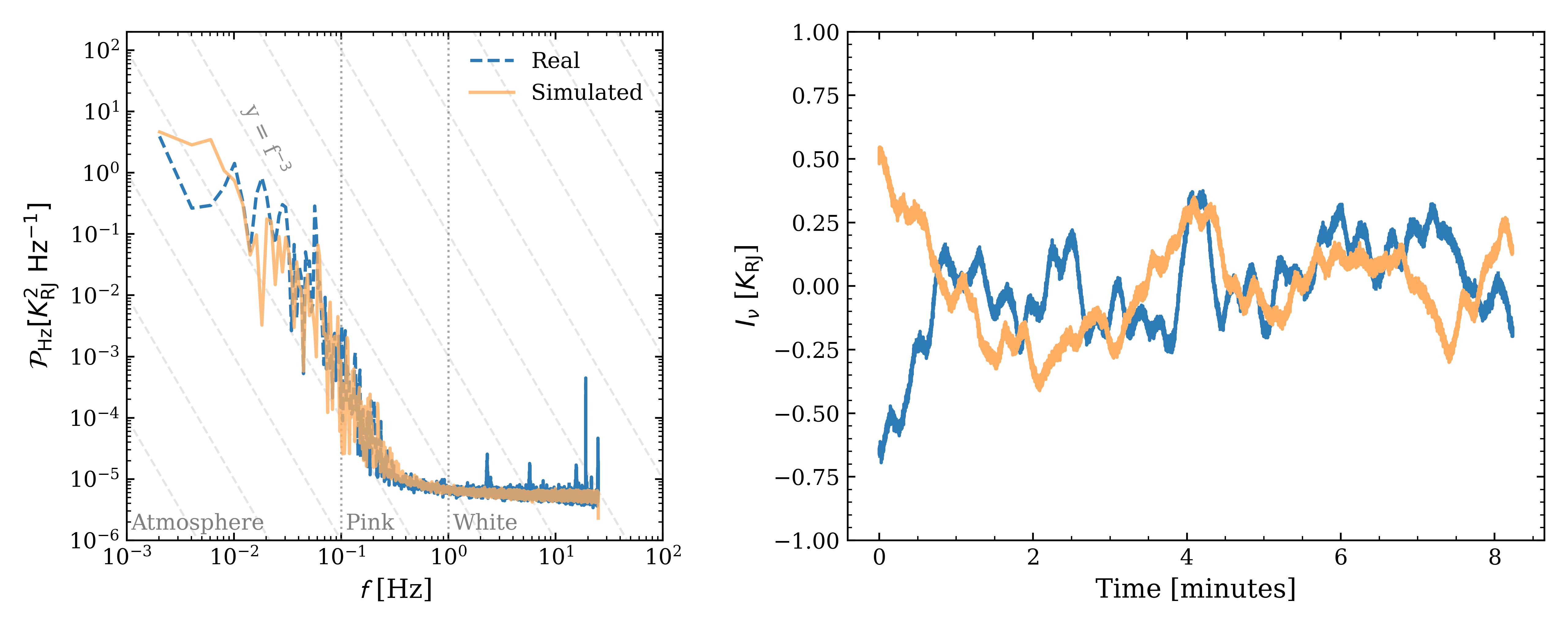}
                    \caption{Comparison between the simulated mock TODs shown in orange with real MUSTANG-2 observations in blue. The weather resembles the atmospheric conditions of the Green Bank Telescope site, as seen in late November around 2 o'clock in the morning, local time. The left panel shows the corresponding mean power spectrum of all detectors. Here, we further highlight the frequency ranges where each noise contribution (atmospheric, pink, and white noise) dominates. The panel on the right shows a single detector's TOD. This figure indicates a remarkable resemblance between real and mock data.
                    }
                    \label{fig:weather}
                \end{figure*}

                In \texttt{maria}, generating an atmosphere necessitates specifying a date and time. \texttt{maria} then generates a random realization of a turbulent 3D atmosphere that evolves in accordance with the given location and time settings.
                Subsequently, the previously defined instrument, telescope, and scanning pattern are used to scan through the evolving atmosphere. This scanning process creates TODs encompassing four variables: the position in Right Ascension (RA) and Declination (Dec), the time stamp, and the brightness temperature of the scan in units of $\rm K_{RJ}$. Furthermore, \texttt{maria} provides the option to overwrite the average PWV of the observation and the corresponding level of PWV fluctuations along the line-of-sight. 
                
    \subsection{Additional Contributions}

        \subsubsection{Detector Noise}\label{sec:noise}
            
            \texttt{maria} provides the option to manually add noise with a power spectrum that scales as $1/f$ (also referred to as `pink' noise) and spectrally flat noise (i.e., scale independent or `white' noise). The additions of pink and white noise are meant to mimic detector and read-out amplifier and gain drift, and thermal and read-out noise, respectively. We note that $1/f$ noise often shares common modes over different detectors. Assuming detectors are completely independent, as done in our implementation, is thus conservative as common modes can often be removed when making maps. 
    
        \subsubsection{Optical efficiencies}\label{sec:efficiency}

            In order to account for the overall response of the detectors to astronomical and atmospheric signals, we included two different optical efficiency parameters for the coupling of the two components. The source of poor optical efficiency is generally dominated by Ruze scattering \citep{Ruze}, followed by a number of efficiency factors that impact the coupling of the instrument to the telescope optics.
            
            In the case of the GBT, the optical efficiency for the main beam at $\nu = 89~$GHz is of order 40\% \citep{Frayer2019, White2022}, while as much as twice that couples back to the atmosphere through scattering terms. To capture both systematics, we add a scalar multiplication to the amplitude of the input astronomical map. We adopt a default value of $\zeta = 0.3$ for coupling of the main beam (i.e., to the astronomical signal). 
    
            For the optical coupling of the atmospheric signal, we set the optical efficiency parameter higher, to $\epsilon = 0.8$, which we assume universally as the default in \texttt{maria}. The reason for a higher optical efficiency here than for the astronomical signal is that light scattered by the telescope, in general, is expected to couple back to the sky. 
            We note that while improving the telescope surface accuracy will improve the coupling factor for the astronomical source, the coupling to the atmosphere should always be higher since it includes the efficiency of the main beam plus that which is Ruze scattered. $\epsilon \approx 0.8$ is derived from the ratio of total optical efficiency (i.e., $\zeta=0.3$) over the Ruze scattering factor ($\approx 40\%$). However, we note that when modeling the atmosphere, the scaling $\epsilon$ is degenerate with the level of PWV fluctuations and its absolute value.

            Finally, we need to consider the atmosphere's opacity, $\tau$. Depending on the elevation of the source, the telescope site, the time of observing, and the observing frequency, the atmosphere absorbs parts of the astronomical background, typically varying from $e^{-\tau} = 0.7-0.9$ at 90~GHz. 
    
            The addition of the optical efficiencies and atmospheric absorption has two impacts. First, the astronomical response is no longer unity, so one must divide the resulting map by that same factor (i.e., $\zeta = 0.3\cdot 0.85$) to ``recalibrate''. Second, it lowers the SNR of the observations by increasing the atmospheric and modeled detector noise (discussed in Sect.~\ref{sec:noise}) at the time stream level with respect to the measured astronomical surface brightness. With all components combined, we can write a simplified antenna temperature equation for \texttt{maria}. This equation summarizes the various components that enter into the simulation and is written up as follows:

            \begin{equation}\label{eq:tsys}
                \rm T_{\rm scan} = \frac{\left(\epsilon~T_{\rm atm} + \zeta~e^{-\tau}T_{\rm astro} + T_{\rm white} + T_{\rm pink}\right)}{\zeta e^{-\tau}}.
            \end{equation}

            \noindent With $\zeta$ and $\epsilon$ the optical efficiency parameters that couple to the atmosphere $\rm T_{\rm atm}$ and the astronomical signal $\rm T_{\rm astro}$, respectively. The resulting $\rm T_{\rm scan}$ is a function of RA, Dec, and time of observing and is given in units of $\rm K_{\rm RJ}$. 
            
    
\section{Validating \texttt{maria} with MUSTANG-2 observations}\label{sec:valid}
    
     Previous work of \citet{morris2022} demonstrated good agreement between reanalysis data and time-ordered data taken by telescopes at the Chajnantor site. However, this comparison has not been extended to other sites like Green Bank. This section compares time streams and noise maps between \texttt{maria} and actual MUSTANG-2 observations to benchmark how accurately \texttt{maria} mimics ground-based large single-dish facilities.

     \subsection{Comparing time streams}\label{sec:tods}

        \begin{figure}[t!]
            \centering
            \includegraphics[width = \hsize]{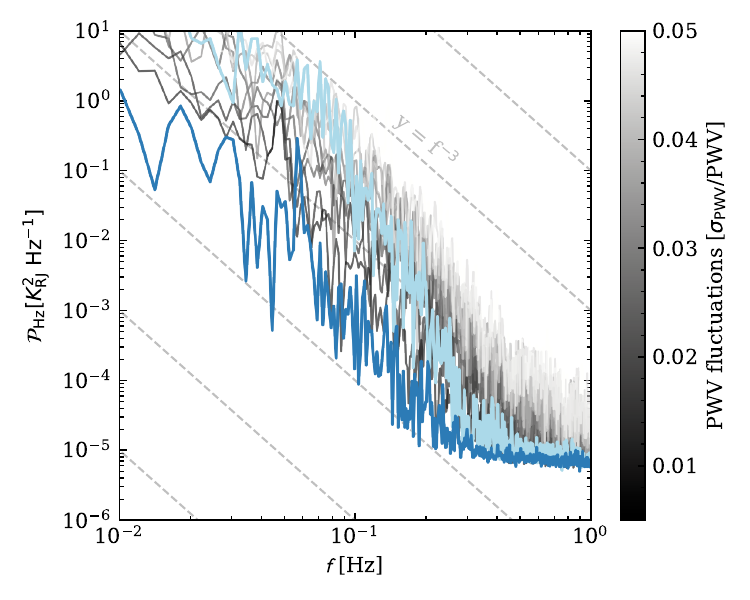}
            \includegraphics[width = \hsize]{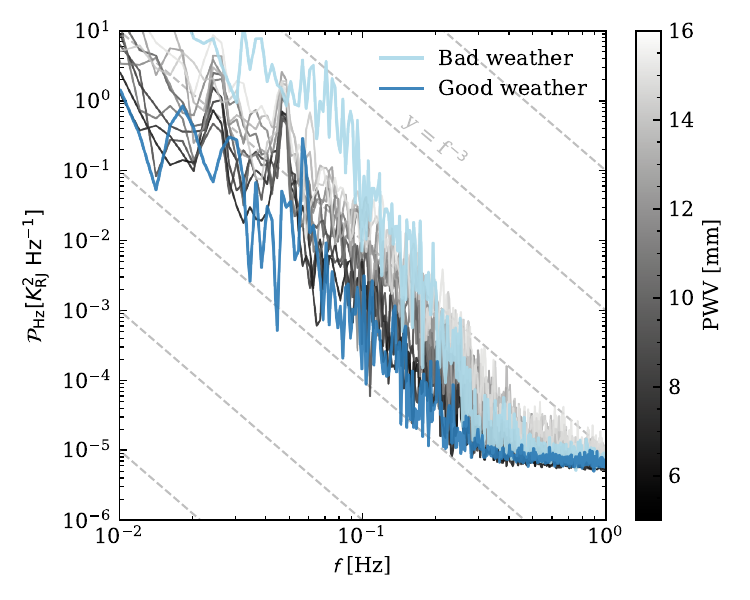}
            \caption{Shape dependence of the atmospheric mean power spectra derived from the synthetic time streams (gray lines) as a function of its simulated weather conditions. These are shown in comparison to real observations indicative of good weather (dark blue, also shown in Figure~\ref{fig:weather}) and worse weather conditions (light blue). The top panel shows the effect of changing the turbulence along the line of sight, ranging from 0.5\% to 5\% of the total amount of PWV (7.7\,mm). The bottom panel shows the behavior of the power spectra with increasing PWV given a constant level of PWV fluctuations of 0.5\% of the total PWV.}
            \label{fig:shape}
        \end{figure}

        \begin{figure}[t]
            \centering
            \includegraphics[width=0.95\hsize]{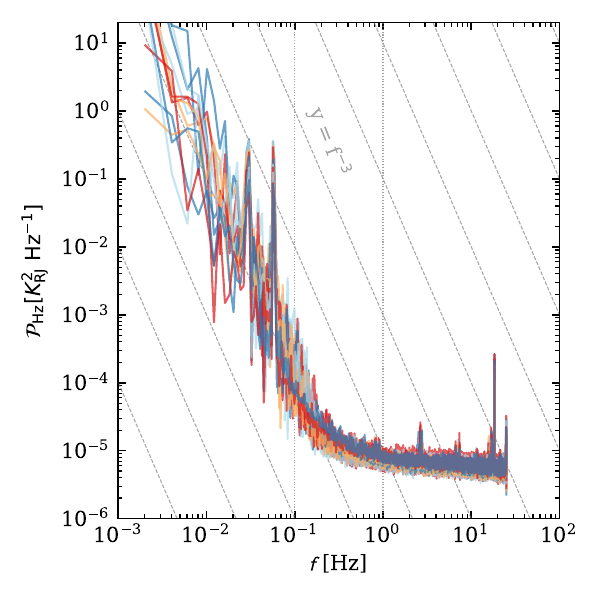}
            \caption{Mean power spectrum of all detectors from 13 consecutive real MUSTANG-2 scans, totaling two hours of on-source time. This figure indicates that the various scans have the same statistical behavior. Hence, a simulated \texttt{maria} time stream matching the statistical behavior of a single scan (Figure~\ref{fig:weather}) will also match that of multiple scans.}
            \label{fig:2hours}
        \end{figure} 
            
        Figure~\ref{fig:weather} compares mock time streams with MUSTANG-2 observations, revealing a close resemblance between the modeled and actual time streams. The time streams have a duration of $\approx 8.5$ minutes, which is the nominal duration of a single GBT/MUSTANG-2 scan. The left panel shows the power spectra of such a scan, which is also depicted on the right. Prior to computing the power spectra, the time series are apodized using a tapered cosine window. The power spectra reveal three different regimes where the atmosphere (left), pink (middle), and white noise (right) are dominant as a function of the sky sampling frequency. We found that a white noise level of T$=1700~\mu$K$\sqrt{\rm s}$ (corrected for the calibration parameter $\zeta$ and the opacity of the atmosphere, see Eq.~\ref{eq:tsys}) was required to reproduce the real power spectrum. The $1/f$-term was also added in order to reproduce the real-time streams. 
        
        Regarding the atmospheric regime in the power spectra, \citet{morris2022} derived that at angles below the outer scale, the power spectra should follow a frequency dependence of $\mathcal{P}\propto\left(f\left[{\rm Hz}\right]\right)^{-8/3}$ when the beams propagating through the atmosphere do not heavily overlap (e.g., in the case of ACT). Conversely, at angles smaller than the near-field width of the beams, the angular atmospheric power spectra should follow a $\mathcal{P}\propto\left(f\left[{\rm Hz}\right]\right)^{-3}$ dependency which is derived from the Fourier transform of the structure-function $C(0)-C(\theta) \propto \theta^2$; see the Appendix in \citealt{morris2022}. In the case of the GBT/MUSTANG-2, the near field lies far above the atmosphere due to the large primary dish size. To illustrate, with the detectors being spaced 1.9$f-\lambda$ apart, the angular separation would be approximately $15\arcsec$ on the sky. Consequently, the two beams would only see a difference of $z\theta \approx 0.2$~m patch of the atmosphere at a distance of 3 km above the telescope while the whole beam is 100~m wide (following the primary dish size), indicating that the beams heavily overlap at the relevant scale heights. Therefore, we would expect the power law behavior of MUSTANG-2 data to follow the $\mathcal{P}\propto\left(f\left[{\rm Hz}\right]\right)^{-3}$ trend as observed in the real data.       

        The shape of the power spectrum is also influenced by both the absolute level of PWV and its fluctuation strength along the line of sight, as illustrated in Figure~\ref{fig:shape}. The top panel shows the power spectrum as a result of varying the turbulence, which is characterized here as the fractional PWV RMS along the line of sight. Here, we froze the PWV value to 7.7~mm. We plot the synthetic TODs along with real MUSTANG-2 observations, one of which is also shown in Figure~\ref{fig:weather}, while the other is another real scan taken on the same day during worse weather conditions. The absolute level of PWV corresponds to the PWV level of the scan taken in worse weather conditions.  In the bottom panel, we varied the overall PWV level while freezing its fluctuations scale to 0.5\% (the level also used to reproduce the real time stream shown in Figure~\ref{fig:weather}). In order to match the atmospheric power at small $f$ for the scan taken in worse weather conditions, Figure~\ref{fig:shape} shows that not the level PWV, but rather a higher level of fluctuations (i.e., turbulence) is needed to reproduce the observations and is thus more indicative of observing bad weather. 
        
        The overall similarity between the mock and true time streams of Figure~\ref{fig:weather} is further confirmed by conducting a $k$-sample Anderson-Darling test between the two time streams depicted in Figure~\ref{fig:weather}. The tests yield a p-value smaller than $p<0.001$. This result strongly indicates a high level of agreement between the real and mock datasets. 
        
        Figure~\ref{fig:2hours} shows the mean power spectra of all detectors from 13 consecutive real 8.6-minute long MUSTANG-2 scans, totaling approximately 2 hours of observation. The alignment of all power spectra indicates consistent statistical properties across scans. Therefore, the comparison between the simulated and real observations presented in Figure~\ref{fig:weather} also applies to longer integration times. 
        
     \subsection{Comparing noise maps}\label{sec:maps}

        \subsubsection{Naive Mapmaking}

             \begin{figure*}[t]
                \centering
                \includegraphics[width = 0.49\hsize, trim={2cm 0 0 0},clip ]{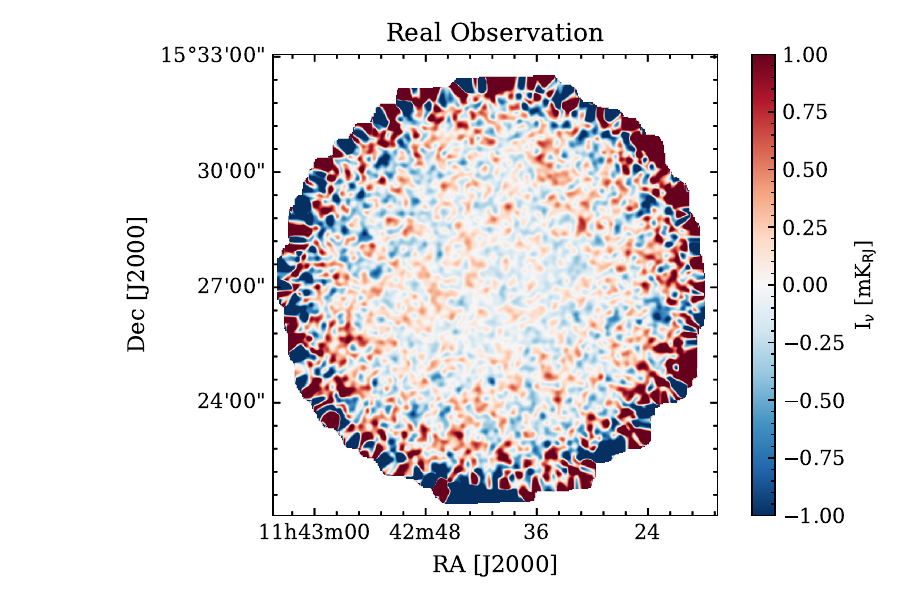}
                \includegraphics[width = 0.49\hsize, trim={2cm 0 0 0},clip ]{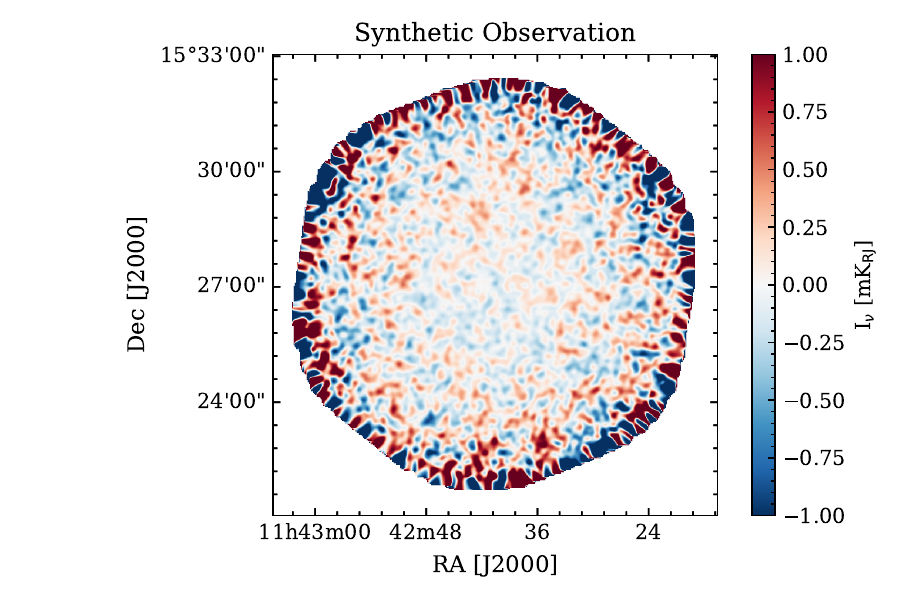}
                \caption{Noise maps, in units of Kelvin Raleigh-Jeans brightness temperature ($K_\textsc{rj}$), for a real (left) and a mock (right) observation. These maps are made using the 8.6~minute long scan shown in Figure~\ref{fig:weather} with the mapmaker implemented in \texttt{maria}. For both the real and synthetic maps, we find an RMS of $148\pm3~{\rm \mu K_{\rm RJ}}$ when estimated within the central $1\arcmin$ region. Here, we corrected the sensitivity of the mock observation with the square root of the ratio of the differences in detectors used. The real observations utilized only 170 detectors out of the 217 available.}
                \label{fig:maps}
            \end{figure*}
            
            Transforming a synthetic time stream into a map of the sky is an art of its own. As \texttt{maria} is designed specifically for generating synthetic time streams, we also implemented a simple, easy-to-use Python-based mapmaking tool into the routine, which we based on MIDAS, the MUSTANG IDL Data Analysis System\footnote{MIDAS when fully unraveled stands for the \textit{MUltiplexed Superconducting Quantum Interface Device Transition Edge Sensor Array at Ninety Gigahertz Interactive Data Language Data Analysis System}.} \citep[see][and references therein]{Romero2020}. But as \texttt{maria} is a generic tool that can simulate various different telescopes and observation strategies, one might not always want to use the same mapmaker. Hence, we also made the TOD exportable to {FITS} tables and Hierarchical Data Format ({HDF5}) files so that one can use whichever mapmaker they prefer.  
            
            Briefly, the mapping tool we developed works as follows: Before binning the TODs into pixels (as also shown on the right panel of Figure~\ref{fig:setup}), we Fourier filter the time streams to remove scales larger than the FoV (in the case for MUSTANG-2 this translates to $f=0.08$) which we assume to be atmosphere dominated. Additionally, in \texttt{maria}, common modes arise solely from the modeled atmosphere (e.g., two adjacent detectors observing the same patch of the atmosphere). Thus, we remove the common mode before mapmaking by subtracting the first mode fit in a singular value decomposition (SVD) performed on the Fourier-filtered time streams. Then, we estimate the average amplitude of the signal measured by all detectors falling within a pixel to create a map. Finally, we smooth the map with the resolution of the telescope (defined as $\theta = 1.22 \lambda/D$). The Fourier-filtered and common-mode subtracted maps for both the real and synthetic time streams are shown in Figure~\ref{fig:maps}, revealing that the resulting noise maps are qualitatively similar. 

        \subsubsection{Limitations of the simulated observations}
    
            Despite the similarity displayed by Figure~\ref{fig:maps}, in this section, we discuss several second-order effects that can still cause minor differences between the real and mock data. The most notable one is that on timescales of several hours, the weather at the Green Bank site might vary, for instance, becoming cloudy. This can worsen or suddenly improve weather conditions, which might not be modeled in \texttt{maria} and lead to variations in the effective beam size. Additionally, surface deformations of MUSTANG-2's primary dish change the effective shape of the beam, which needs to be accurately calibrated, as well as inaccuracies in the bandpass, which are not modeled in full detail in \texttt{maria}. Further, the scanning strategy \texttt{maria} adopts is not exactly the same as that used by the GBT in MUSTANG-2 observations, as the real one features a less regular scanning trajectory. The real observations also include contamination in the form of spikes at high sampling frequencies ($f~[\rm Hz]>2$) that could originate from the pulse tube or other vibrating objects mounted on the GBT, such as the feed arm.
    
            Noteworthy as well; our model for the $(1/f)$ noise components is likely an oversimplification, as correlations in the instrumental $(1/f)$ noise are not fully accounted for in the \texttt{maria} simulations. Further, the MUSTANG-2 detectors are not background-limited, meaning that we had to manually adjust the level of white noise in the simulations to match the observations in Figure~\ref{fig:weather}. Although \texttt{maria} has the option to use background-limited $(1/f)$ estimates of the white noise based on the implementation of \citet{Bryan2018}, a better characterization of which part of the optical/signal path (e.g., amplifiers, readouts, detector thermal drifts, optical loading due to spillover, or surface deformations) causes which noise contribution in the noise power spectra could help improve the noise model when making maps. 

            Finally, although the atmospheric component in \texttt{maria} is state-of-the-art, it has its limitations in accurately creating atmospheric corruption. \texttt{maria} adopts average quantities for typical minute-to-minute variations in PWV and the absolute scale height, which might vary over time \citep[see, e.g.,][for a more detailed discussion]{Morris2024}. Further, there are unknowns, such as how the size of the boundary layer affects inhomogeneities in the atmosphere at different heights, how the local terrain influences local turbulence, and the effect of a non-constant horizontal wind velocity. However, as stated at the beginning of this section, these adjustments only have minor impact on the overall simulated atmosphere, as we show in Figures~\ref{fig:weather},~\ref{fig:shape}, and~\ref{fig:maps}, we can accurately retrieve the statistical behavior of the observations and obtain quantitatively similar noise maps.
            
            \begin{figure}[t]
                \centering
                \includegraphics[width = \hsize,trim={1.2cm 0 0.2cm 0},clip]{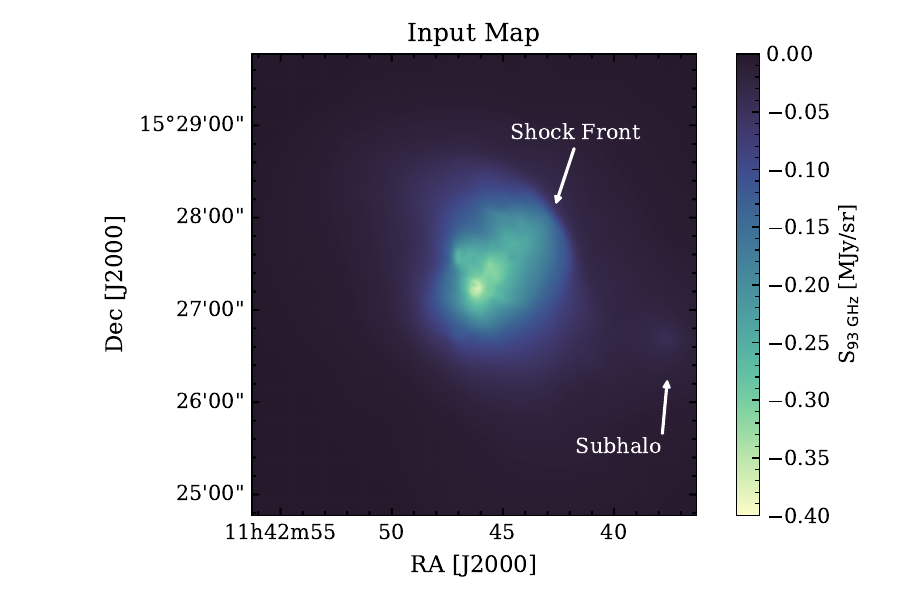}
                \includegraphics[width = 0.96\hsize,trim={1.2cm 0 0.2cm 0},clip]{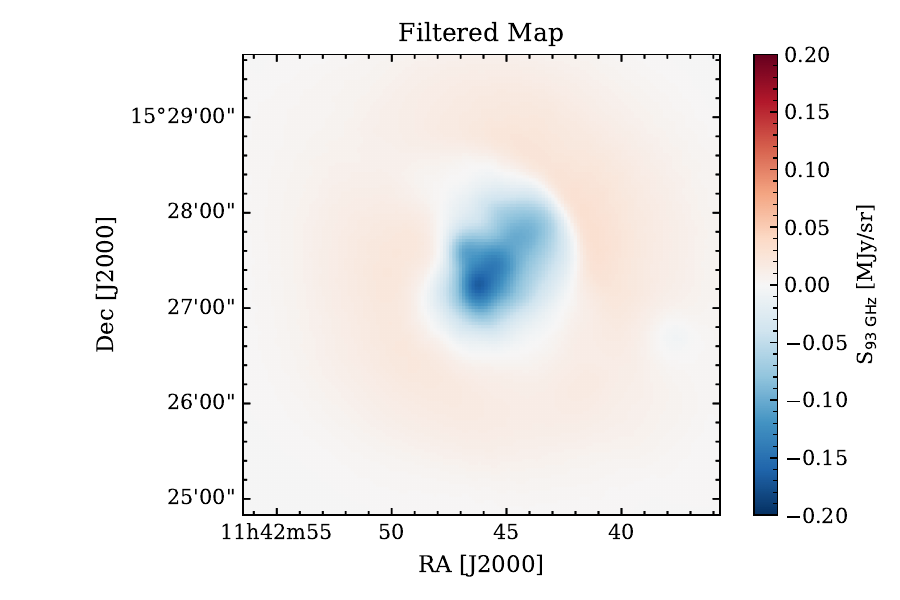}
                \includegraphics[width = 0.96\hsize,trim={1.4cm 0 0.2cm 0},clip]{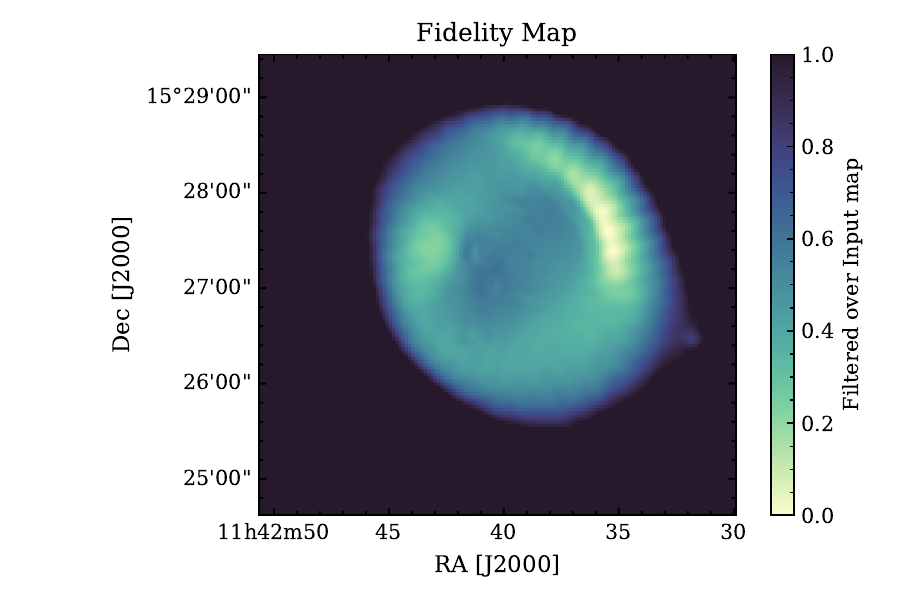}
                \caption{Synthetic images of the thermal SZ effect in units of MJy sr$^{-1}$. From top to bottom, we show the input map, the filtered noiseless map obtained from \texttt{maria}, and the fidelity map (filtered over the input map). This figure illustrates the effect the telescope has on the surface brightness distribution of extended signal, even without adding noise to the data. Here, we saturated the image where the ratio exceeds 1.}
                \label{fig:mock-observations}
            \end{figure}
        
            \begin{figure}[t]
                \centering
                \includegraphics[width = \hsize,trim={1.2cm 0 0.05cm 0},clip]{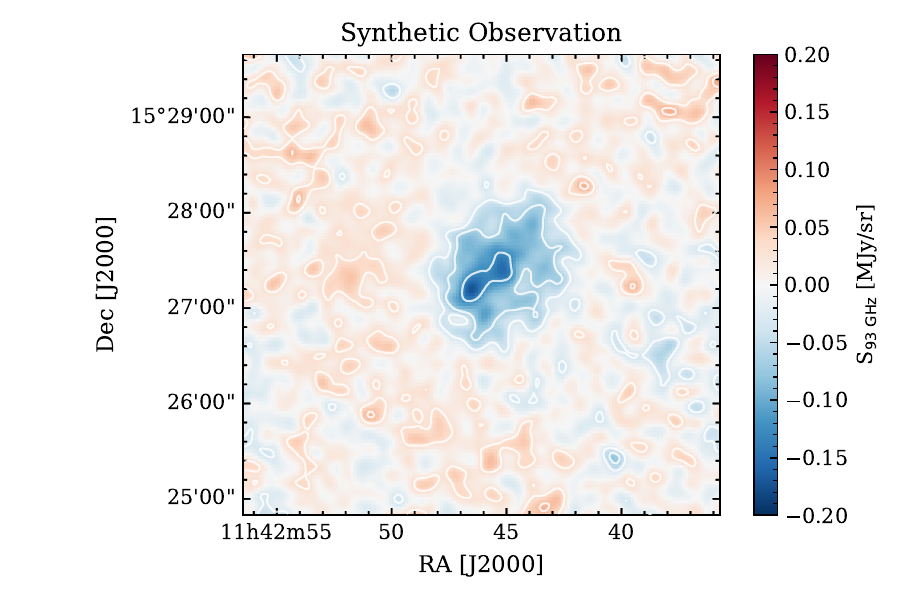}
                \caption{45-minute long mock observation of the thermal SZ effect as seen by MUSTANG-2 on the Green Bank Telescope generated with \texttt{maria}. The mock observation includes atmospheric and $1/f$ (pink) noise. The time streams are Fourier filtered and common-mode subtracted before imaged. The contours are drawn at $[-8, -4, -2, 2, 4]-\sigma$ with $\sigma$ obtained from a circular aperture from the central region in a source-free run.}
                \label{fig:noisy_mock}
            \end{figure}
            
    \subsection{Forecasting MUSTANG-2 observations}\label{sec:M2forcast}
        
        As described in Section~\ref{sec:method}, \texttt{maria} offers the option to scan over a celestial object while traversing the atmosphere and as the noise properties \texttt{maria} derives are close to reality (as described above), we can consequently make mock observations for MUSTANG-2, akin to how \textsc{casa} \texttt{simobserve} works for interferometric observations. To illustrate, we use a map of the thermal SZ effect from a galaxy cluster extracted from the ``Dianoga'' hydrodynamical zoom-in simulations \citep{Rasia015}. Following the methodology employed by \citet{DiMascolo2023}, we transformed the particle information into a projected high-resolution Compton-$y$ map of the hot gas within the intracluster medium (ICM) of a galaxy cluster. An example of such a map is shown on the top panel of Figure~\ref{fig:mock-observations}. This panel shows a cluster at $z = 1.0$ with a halo mass of $M_{500,c} = 8\times 10^{14}~\rm M_\odot$. Here, we define $M_{500,c}$ as the mass of the system within a radius having an average density of 500$\times$ the critical density of the Universe at that redshift. This cluster is undergoing a merger, showing features due to merger-driven shocks in the north and an additional faint subhalo component to the west of the main cluster. 
        
        To make mock MUSTANG-2 observations from this cluster, we used our mapmaking tool following the exact same procedure as described in section~\ref{sec:maps}. We first ran \texttt{maria} without white, $1/f$, and atmospheric noise to show the effects the scanning, spatial resolution, filtering, processing, and common-mode subtraction have on the input map. The results of the ``filtered'' map are shown in the middle panel in Figure~\ref{fig:mock-observations}. Then, the ratio of the filtered map over the input map is shown in the bottom panel of Figure~\ref{fig:mock-observations}, which we will refer to as the fidelity map. We smoothed the input with the diffraction-limited beam before computing the fidelity map to prevent beam-smearing effects. We also shifted the flux of the filtered map so that the maximum pixel has a flux value of 0 to correct for any dc-offset between the two maps. We saturated the image where the ratio exceeds 1. 
        
        The common-mode subtraction and Fourier filtering cause the peak flux to drop by a factor $\approx0.5$. The fidelity map illustrates the difficulty in retrieving the astronomical signal. The two features described previously, namely the distinct shock front and the fainter subhalo, are visible in the fidelity map, meaning that they are suppressed in the filtered map and consequently in the observation itself. 
        A peak flux reduction of $\approx50\%$ can significantly impact the possibility of detection, as also seen in Figure~\ref{fig:noisy_mock}. Figure~\ref{fig:noisy_mock} shows the full mock observation of the cluster in which we add the atmosphere, pink, and white noise to the time streams. This 45-minute long execution block clearly detects the tSZ effect with a peak SNR of $-8\sigma$; however, the shock front and the more extended, faint features of the cluster go undetected. 

\begin{figure}[t]
                \centering
                \includegraphics[width = 0.99\hsize]{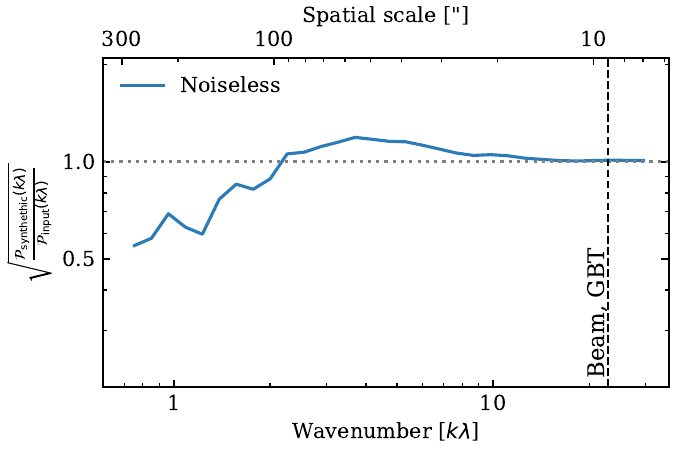}
                \caption{Transfer function 
                of the mock MUSTANG-2 observation, derived from the power spectra of the filtered map shown in Figure~\ref{fig:mock-observations}.
                The transfer function is normalized by that of the input map shown in Figure~\ref{fig:mock-observations} and smoothed with a top hat function.}
                \label{fig:tranfer_M2}
            \end{figure}

        To further quantify the scale dependence of how information is lost due to filtering and processing, we compute the transfer function of the filtered map with respect to the input map (see Figure~\ref{fig:tranfer_M2}). Here, the transfer function is defined by taking the square root of the power spectra of the mock observation and dividing it by the power spectra of the input map (which is corrected for beam smearing). To prevent the power spectra from blowing up at large wave numbers (smaller scales), we added white noise to the image at a level of a tenth of the measured sensitivity in the central region in Figure~\ref{fig:noisy_mock}. The power spectra are estimated by binning the Fourier transform of the map in log-spaced bins from $0~{\rm k\lambda} < k < 40~{\rm k\lambda}$ and taking the average of the amplitude of the Fourier transform per $k$-bin. The ``noiseless'' curve shows the ideal transfer function for the observations. It exhibits a flat $k$-dependence up to $\approx 30\arcsec$, after which it drops to $\approx 0.5$ by $\approx 300\arcsec$, which is similar to what is derived from the fidelity map shown in Figure~\ref{fig:mock-observations} and agrees well with real MUSTANG-2 observations \citep[see e.g.,][]{Romero2020}.

        However, we note that the transfer function drops partially because of the Fourier filter applied to the streams before making maps. More advanced mapmakers, which adopt a maximum likelihood mapmaking approach, such as the mapmaker used with ACT observations \citep{Dunner2013} and \texttt{}{minkasi} (Sievers et al., in prep), have indicated a more flat transfer function, e.g., being better equipped to recover larger angular modes at converged scales \citep[see, e.g.,][]{Romero2020}. Nevertheless, Figure~\ref{fig:tranfer_M2} shows that regardless of the noise in the observations, scanning over a source reduces the recovered signal, and hence, the real observations must be corrected for such a transfer function. In future work, we will go into more depth on how well one can recover the input given the synthetic time streams. We will test current ``observational'' tools on how to retrieve radial surface brightness profiles and generic properties such as the halo mass of the cluster, as well as the feasibility of measuring features such as subhaloes and shock fronts. In this work, we focus on illustrating \texttt{maria} and how we can use it to create synthetic observations to study the feasibility of current, as illustrated in this section, and future facilities. We will further expand upon the latter in the next section. 
        
\section{AtLAST forecasts}\label{sec:AtLAST}

        To illustrate that \texttt{maria} can be used to evaluate science cases for future facilities, we provide a section on adjusting the simulation class for an AtLAST-like observation done with a relatively small-scale bolometer array design. We will also briefly discuss the performance of such a telescope in comparison with other single-dish facilities. In particular, we discuss the noise properties of the time streams and their level of common modes (see Section~\ref{sec:large_scales}) as well as the sensitivity at beam scales (see Section~\ref{sec:noise_smallscales}). We will also make synthetic images of a galaxy cluster that comes from the Diagona hydrodynamical simulations for this simulated instrument and compare them and the resulting transfer function with that of MUSTANG-2 (see Section~\ref{sec:forecast_atlast}).

        \subsection{Simulation set up}\label{sec:AtLAST_setup}

            AtLAST is a proposed 50-meter diameter single-dish facility to be built in the Atacama desert near ALMA. AtLAST has already undergone a 4-year EU-funded design study and has been selected for a new EU infrastructure development grant starting in 2025. AtLAST's concept is that of a facility observatory hosting up to six massive instruments and serving a wide range of science cases. For the purpose of this work, we focus on a particular observing mode, class of instrumentation, and science goal, but we note that future works will explore AtLAST's observing capabilities more broadly as well as make forecasts for its scientific applicability. We note here that the telescope will have a 4.7-meter focal plane, corresponding to an instantaneous field of view of 2$^\circ$, and will allow extremely fast scan velocities of up to $3^\circ \,\rm s^{-1}$ (e.g., in the middle of an azimuthal scan), and an acceleration of up to $1^\circ \,\rm s^{-2}$ as defined in \cite{Klaassen2020}.

            \begin{figure}[t]
                \centering
                \hspace{-5mm}\includegraphics[width = 0.9\hsize]{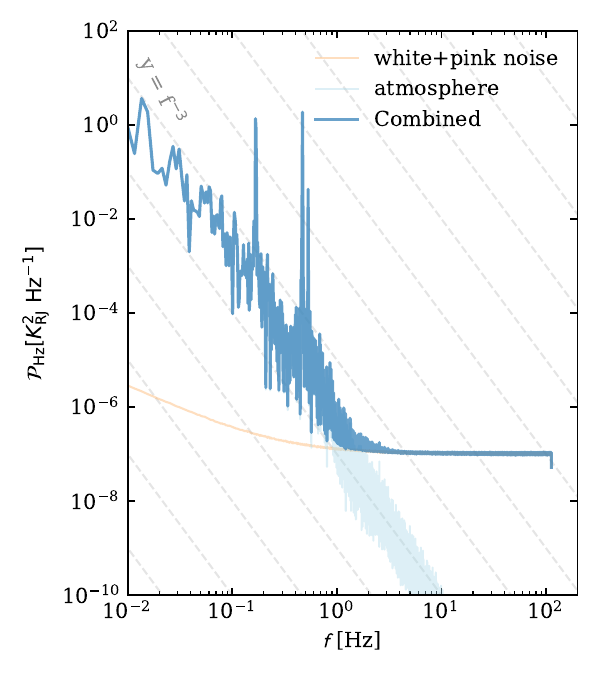}
                \includegraphics[width = \hsize,trim={1.2cm 0 0.5cm 0},clip]{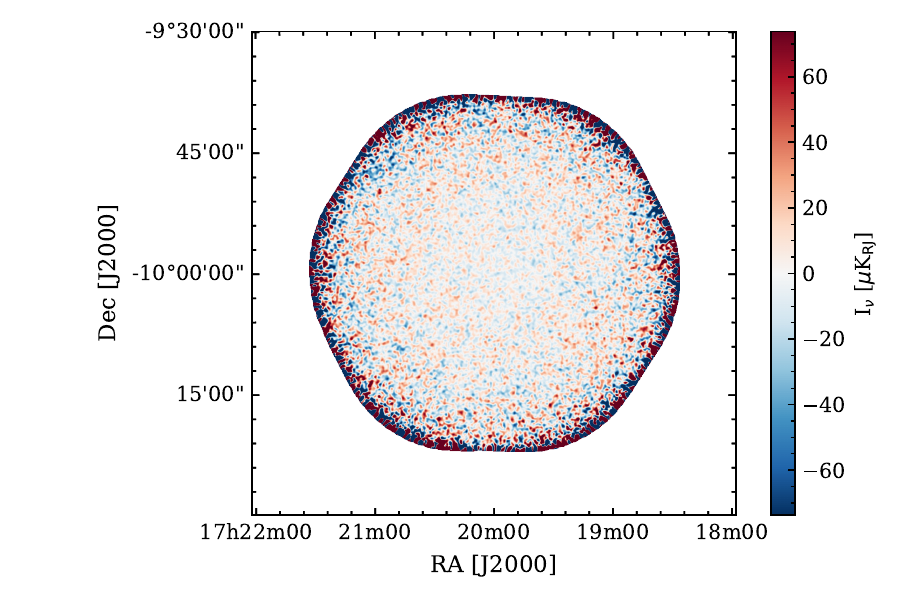}
                \caption{A simulated AtLAST noise power spectrum (upper panel) and the noise map (lower panel). The simulations are run assuming a broadband monochromatic imaging array centered at 93 GHz with 3000 detectors with a $0.25^\circ$ instantaneous FoV. The time streams are common mode subtracted and Fourier filtered before imaged. More details of the observational setup can be found in Section~\ref{sec:AtLAST}. The resulting RMS of the noise map is 12.0 $\mu$Jy/Beam which is equivalent to 12.2 $\rm\mu K_{\rm RJ}$.}
                \label{fig:AtLAST_noise_maps}
            \end{figure}
            
            To Nyquist sample the beam, which has a FWHM $\theta = 1.22 \lambda/D \approx16\arcsec$ for a 50-meter telescope at 93~GHz and assuming an average scanning velocity of $1^\circ$/s, we need to scan with a read-out rate 90$\times$ that of MUSTANG-2, resulting in a sampling of 900 Hz. AtLAST will also observe at high frequencies up to $\nu \approx 950~$GHz, increasing the rate to $\approx9$~kHz. 
            Furthermore, to maintain the same sky sampling, the number of detectors increases with the frequency squared if the FoV size remains constant. For the 93~GHz band, we need $\approx200,000$ detectors to fill the $2^\circ$ FoV with the detectors placed at 1$f-\lambda$. Hence, for the high-frequency ($\nu = 900~$GHz), AtLAST would need 20 million detectors to fill the FoV. Fortunately, as noted in \cite{Klaassen2020}, the number of detectors increases by roughly a factor of ten every seven years, implying that instruments comprising of order 1 million detectors could be expected by 2030. However, both the high sampling rate and the number of detectors in the array result in stupendously large data rates. Simulating this with \texttt{maria} is thus computationally expensive and requires a large working memory ($>1$TB). The MUSTANG-2 mock observations finished its run within five minutes on a 2019 MacBook Pro with a 2.6 GHz 6-Core Intel Core i7 processor, but with the current implementation of \texttt{maria} (v1.0), an even bigger server (e.g., a 64-core workstation with 1~TB of random access memory) is not enough to mimic AtLAST-like configurations. 

            To temporarily overcome the data rate problem, we simulate here an instrument with a reduced FoV of only $0.25^\circ$ ($15\arcmin$) and limit the maximum scanning velocity to $0.5^\circ \, \rm s^{-1}$, which occurs at the center of the daisy scan. This will limit the number of detectors to 3000 (a factor of 32 smaller than AtLAST could host) and lower the data rate by sampling at 225 Hz (12 times lower). These adjustments make the data rates feasible to run with \texttt{maria} as we continue to improve the efficiency of the code. For the remainder of this work, the observations are set to the following configuration: it will point at a source in the position of an RA and Dec of $(260.0^\circ, -10.0^\circ)$ at the date of 2022-08-01T23:00:00 which results in an average elevation of 57.3$^\circ$. The default scanning radius of the daisy scan will be the size of the FoV. Furthermore, we will simulate one band for all detectors centered on 92~GHz with a bandwidth of 52~GHz, which is ideal for detecting the SZ effect at those frequencies. Finally, we will simulate the atmosphere to a scale height of 3000 meters above Chajnantor and use an RMS fractional variation in PWV of 5\% to get a total temperature of around $T_{\rm scan}\approx 28~K_{\rm RJ}$ which is a typical atmospheric value \citep{982447} when observing at the APEX, ALMA, or future AtLAST site.

            Further, we adopt the measured mean sensitivity of the PA6 array ($\rm T=266~\mu K~\sqrt{\rm s}$ at $\nu = 90$~GHz, Naess et al. in prep.) of ACT as a white noise level for each detector used in our setup. As the ACT detector noise performance is nearly background-limited, better sensitivities per detector per time stamp cannot easily be achieved. Further, we normalized the ACT sensitivity with the ratio of the different bandpasses and corrected for the optical efficiency we simulated; Here, we assume similar optical efficiencies as for the GBT and MUSTANG-2. The resulting white noise level is four times lower than that of MUSTANG-2. 
            Then, we scaled the pink noise level down from what we used in the MUSTANG-2 case with the same amplitude as the difference in the white noise levels.
            All this results in an antenna temperature of around $T_{\rm scan}\approx 72~K_{\rm RJ}$ after correcting the time streams for the optical efficiency and the atmosphere's opacity. The resulting power spectrum and noise map of an 8.6-minute long scan are shown in Figure~\ref{fig:AtLAST_noise_maps}.\footnote{Adopting an 8.6-minute scan duration for AtLAST is arbitrary and was chosen solely for consistency with the MUSTANG-2 comparison. The scan duration for AtLAST is not fixed and will likely depend on the specific instrument and scientific goals.}
            Before making this map, we highpass filter the time streams in Fourier space to remove scales larger than the field of view (i.e., $\lesssim0.5$~Hz in the time domain, depending on the scanning speed). We then remove the first principal component, as is often done with real observations. The Fourier filtering removes the peaks caused by the scanning harmonics seen below 0.5~Hz in the upper panel of Figure~\ref{fig:AtLAST_noise_maps}. The Fourier filtering also removes most of the atmospheric contributions from the time streams.                 

    \subsection{Spatial scale considerations}\label{sec:atlast_noise}

        \begin{figure}[t]
            \centering
            \includegraphics[width=\hsize]{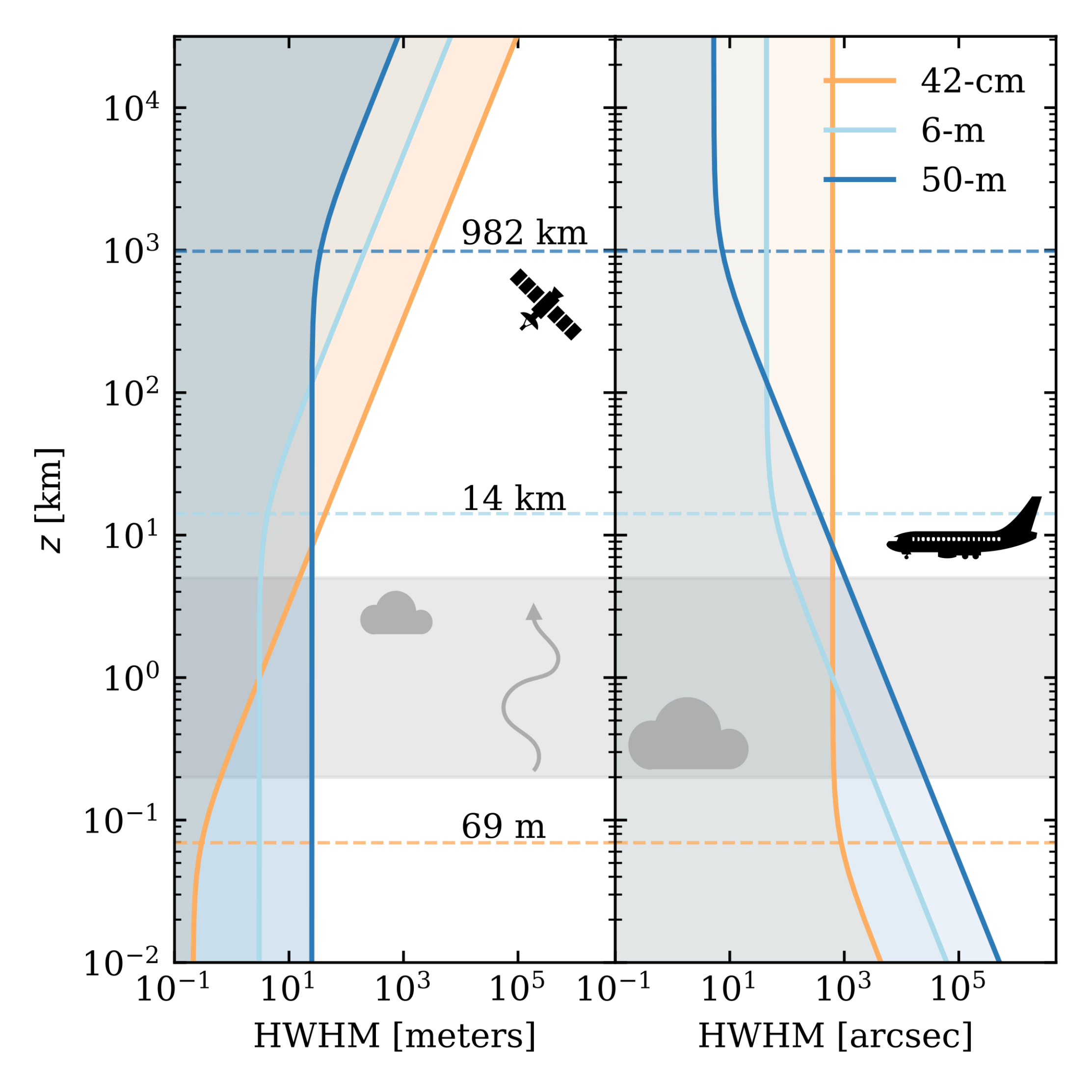}
            \caption{This figure illustrates the geometry of how beams propagate through the atmosphere at a height $z$, perpendicular to the ground following Eqs.~(\ref{eq:beam},\ref{eq:near}). The left panel shows the propagation in physical units (m), and the right panel shows it in angles (arcsec). We show three different beams corresponding to three different apertures, namely 42-cm (yellow), 6-m (light blue), and 50-m (dark blue). The Rayleigh heights of these beams are given at 150~GHz. In light gray, we show the heights where \texttt{maria} simulates the atmosphere ($z=0.2-5$~km).
            For reference, we show the typical heights relative to Llano de Chajnantor of cumulus clouds, a passenger jet at cruising altitude, and the orbital radius of the International Space Station.}
            \label{fig:beams}
        \end{figure}

        When characterizing the noise properties of a telescope, two regimes need consideration. The first is regarding the recovery of large angular scales of the sky (low $\ell$), crucial for applications such as CMB observations and line intensity mapping (LIM) experiments. The second regime involves the sensitivities at beam scales (high $\ell$), which is essential for tasks like eliminating contaminating sources such as the cosmic infrared background or conducting targeted studies of, for instance, the circumgalactic medium in distant galaxies. In this section, we will briefly address both regimes.

        \subsubsection{Large spatial scales}\label{sec:large_scales}

            In addition to collecting area and resolution, one of the advantages of constructing a large ($\geq 30-$meter) single-dish telescope, such as AtLAST or CMB-HD \citep{CMBHD2019, CMB-HD2022}, turns out to be the enhanced correlation of atmospheric effects across beams of adjacent detectors, which facilitates better removal of the atmospheric signal. 
            
            In \texttt{maria}, we assume diffraction-limited beams, with a FWHM Gaussian beam defined as 
    
            \begin{equation}\label{eq:beam}
                {\rm FWHM}~(z) = {\rm D} \sqrt{z^{-2} + z_r^{-2}}~\left[\rm rad.\right],
            \end{equation}

            \noindent as a function of distance $z$ perpendicular to the surface. Here, ${\rm D}$ is the size of the primary dish, and $z_r$ is the Rayleigh height in meters which is defined as
            
            \begin{equation}\label{eq:near}
                z_r = \pi~\frac{({\rm D}/2)^2 }{\lambda}~,
            \end{equation}
            
            \noindent where $\lambda$ is the wavelength of the observation.
            
            The Rayleigh height sets the boundary between the near and the far-field. From Eq.~\eqref{eq:near}, we infer that in physical units (m), the near field beam behaves as a cylindrical column that transverses through the sky and diverges only at higher altitudes for larger dish sizes, as illustrated in Figure~\ref{fig:beams}.
            Based on the geometry of the beam, we infer that a large dish sees more of the same atmosphere between two beams than a 6-m or a 42-cm aperture, which are the apertures sizes of the large aperture telescope (LAT) and small aperture telescope (SAT) used in SO. To illustrate, at a scale height of 1 km and adjacent detectors at 150~GHz spaced $1 f-\lambda$ apart, with AtLAST, one detector's beam would be shifted only by 48~mm relative to the next adjacent beam (i.e. roughly one-thousandth of the 50~meter column of atmosphere probed in the near field). This implies that one detector sees 99.1\% of the same of the atmosphere as its neighbor. For the 6-m apertures, the same considerations imply the shift would be 87.9\%, making time-lagged atmospheric signal removal more difficult. For the 42-cm, this would be even worse; however, as Figure~\ref{fig:beams} shows, a 42-cm aperture results in an extremely short near field, with the beams beginning to diverge around 69 meters, thus increasing the overlap in the instantaneous volumes of the atmosphere probed, which also should facilitate better atmospheric removal in post-processing. 
            

            \begin{figure}[t]
                \centering
                \includegraphics[width = \hsize]{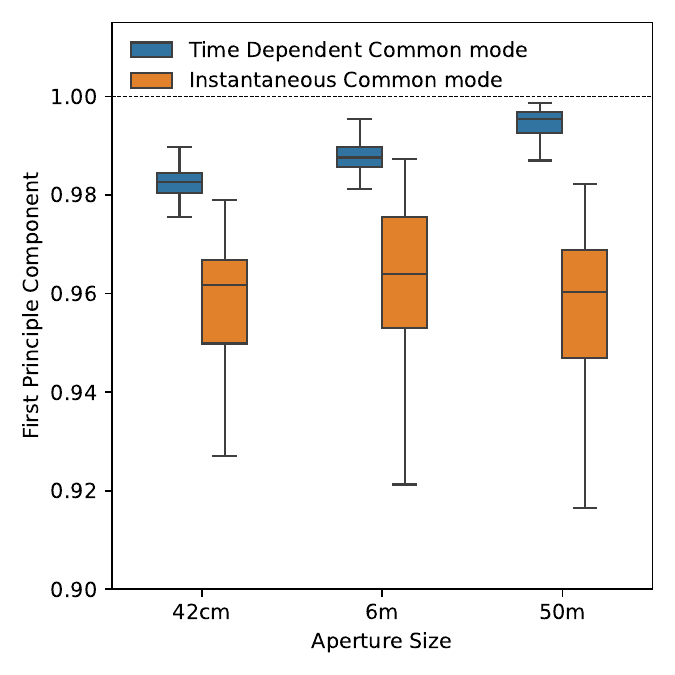}
                \caption{The estimated cumulative contribution rate of the eigenvalue of the first principal component of a time stream containing only atmospheric contributions simulated with a 42-cm, 6-m, and 50-m sized aperture. Here, we show the level of correlation between all the detectors for the whole $2^\circ$ FoV for half a second of integration time (referred to as the Instantaneous Common mode) and for two neighboring detectors over a 60-second integration period (referred to as the Time Dependent Common mode). 
                This figure indicates that the level of the correlated atmosphere is highest for the 50-m sized aperture. 
                }
                \label{fig:PCA}
            \end{figure}    

            With \texttt{maria}, we can test the trade-off between having a short near field with quickly diverging beams versus a more extended near field probing a larger column of common atmospheric signals. To quantify the observed degree of correlated atmosphere for different types of aperture sizes (42-cm, 6-m, and 50-m), we compare various eigenvalues of the first principal component using a principal component analysis (PCA) on time streams simulated with only atmospheric contributions. Since PCA is not scale-invariant, we standardized the time streams of various mock observations before estimating the eigenvectors. 
            When setting up the arrays in \texttt{maria}, we placed 3000 detectors in a $2^\circ$ FoV for all three aperture sizes. 
            Therefore, the detectors in the array that correspond to an aperture size of 50-m will be spaced farther from each other in terms of $f-\lambda$ than the 6-m and 42-cm apertures because of the different beam sizes. Due to this sparseness, both the 50-m and 6-m cases have detectors in their array spaced beyond $2~f-\lambda$, but the beams overlap for the 42-cm aperture. 
            The detectors were then set up to stare through the atmosphere for a period of 60 seconds at the Llano de Chajnantor site. This experiment is thus set up similarly to the example outlined above.
            
            Figure~\ref{fig:PCA} shows a boxplot of the cumulative contribution rate of the first eigenvalues from 50 different atmospheric runs for two types of common modes: the time-lagged common mode and the instantaneous one.
            The instantaneous component measures the degree of common modes for all detectors in the $2^\circ$ FoV integrated over half a second of integration time, which, as outlined above, should be larger for the 42-cm apertures. However, this is not significantly measured, as shown in  Figure~\ref{fig:PCA}. 
            This could be due to the low level of turbulence in the atmosphere at the Chajnantor site. In that case, even for non-overlapping beams, the observed atmosphere is still heavily correlated over the $2^\circ$ FoV, thus reducing the advantage of overlapping beams for small apertures. 

            Regarding the time-lagged common mode, we performed the PCA on two neighboring detectors for all simulated time steps (1 minute of simulated TODs). From the back-of-the-envelope estimate above, the 50~m dish should detect more correlated atmosphere when it transverses through the cylindrical column than the smaller-sized apertures. The results shown in Figure~\ref{fig:PCA} indicate that the 6~m and 42~cm scenarios perform less well than the 50~m dish by $1.6\sigma$ and $2.5\sigma$, respectively, which is in line with the calculation outlined above. This suggests that the 50~m dish telescope has the highest level of measured common mode signal in the time streams. 

            A fundamental assumption here is that a higher level of common mode signal should allow for better mitigation of the atmosphere, leading to the recovery of lower $\ell$. This simulation shows that a 50~m dish could better recover signals at lower $\ell$ than 6~m or 42-cm sized apertures for a similar FoV. ACT has already demonstrated with a 6~meter aperture and 1$^\circ$ FoV the ability to recover scales down to $\ell \approx 350$ \citep[see, e.g.,][]{Coulton2024, Qu2024} and SPT probing down to $\ell\approx50$ \citep{Henning2018}. Therefore, even if one assumes that the spatial scales one can recover are limited to scales less than or equal to the size of the FoV, this implies that AtLAST could also carry out CMB observations on similar or larger angular scales than the current CMB experiments.

            We note, however, that the comparison here limits all three aperture-sized telescopes with a FoV of $2^\circ$, while the SATs in SO will have $\approx30^\circ$ FoV, which implies the recovery of even lower $\ell$, although the atmosphere at these scales will not be as correlated. Regardless, this experiment highlights the need for techniques to mitigate atmospheric effects, both in hardware (for instance, by utilizing half wave plates) and in software (such as to improve upon the entire TOD to map pipeline). Additionally, the trade-off of having a large $>30$~m single dish telescope is to be also sensitive to high-$\ell$.   
                
        \subsubsection{Small spatial scales}\label{sec:noise_smallscales}
        
            \begin{figure*}[t!]
                \includegraphics[width=\textwidth]{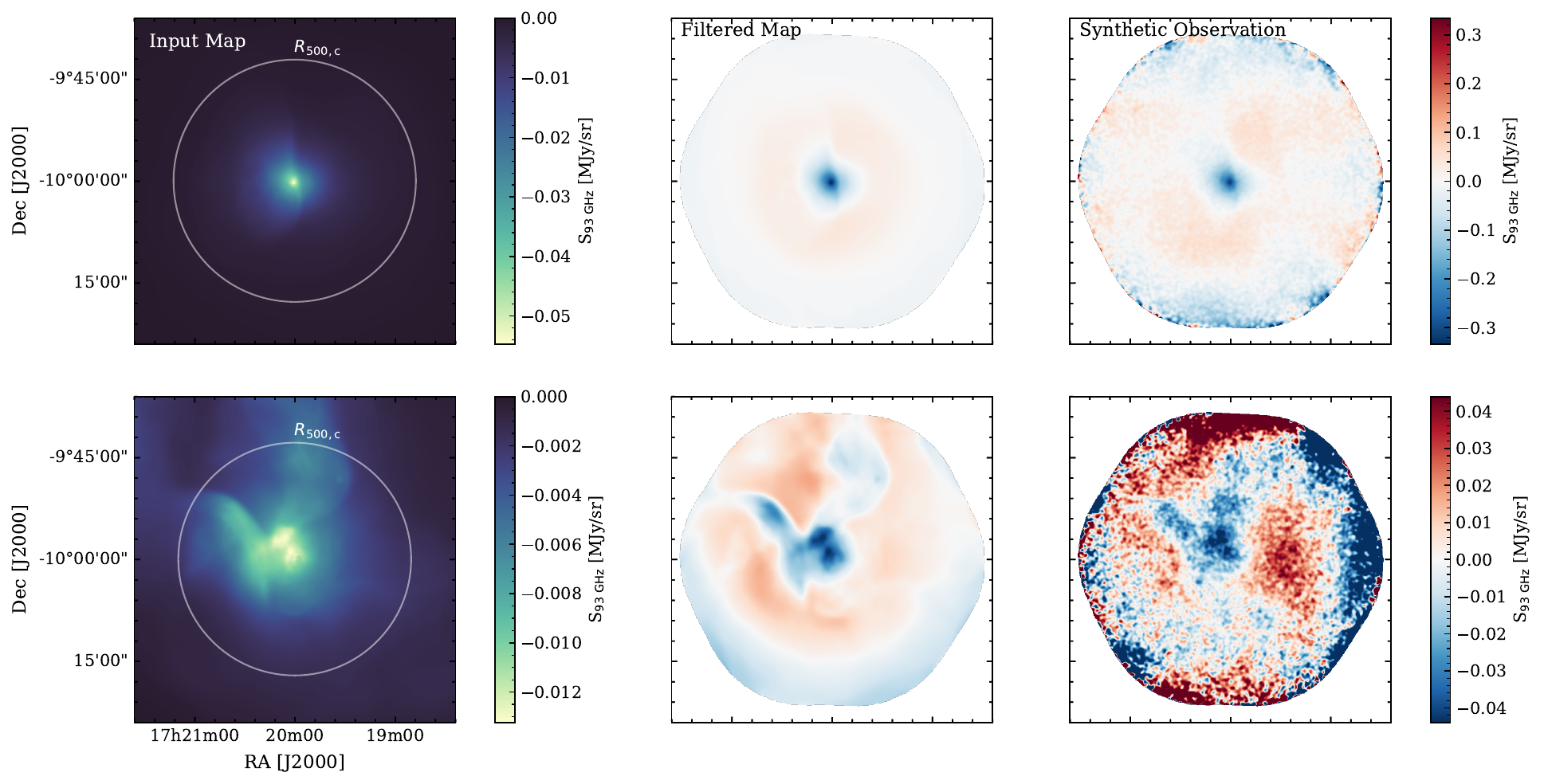}
                \caption{Simulated low redshift ($z = 0.0688$) clusters of galaxies, with masses $M_{\rm 500,c} = (8.8,\ 7.8)\times 10^{14}~{\rm M_\odot}$ (top to bottom), from the Dianoga suist of hydrodynamical simulations are shown on the left. The middle column shows an uncorrupted \texttt{maria} simulation of those clusters, highlighting the effects of an AtLAST-like transfer function. However, instead of simulating the complete $2^\circ$ FoV, we mimic an instrument that only utilizes a $0.25^\circ$ instantaneous FoV to facilitate simulation. The right column shows simulated observations, where atmospheric, white, and pink noise appropriate for 43 minutes of on-source integration time is added.}
                \label{fig:AtLAST_mock-obs}
            \end{figure*}

            Regarding the sensitivity at beam scales (high $\ell$), we can compare \texttt{maria} simulations with the predictions from the AtLAST sensitivity calculator.\footnote{See \url{https://senscalc.atlast.uio.no/}} There, the point source sensitivity is estimated by using the \texttt{CASA} corruption model set to \textit{Total Power mode} \citep{CASA2022} and scaling up the diameter of a single ALMA antenna to 50 meters. The sensitivity calculator is appropriate for maps of the line emission using state-of-the-art heterodyne instruments, which trade off a worse noise performance with respect to a bolometer to preserve phase coherence and achieve much higher spectral resolution ($R>10^6$). However, this model does not account for correlated atmosphere, nor are scanning patterns implemented. Hence, using the sensitivity calculator to make forecasts for bolometric (e.g., broadband continuum) instruments is not ideal as it will act as if it has a single detector on the sky that will stare at the celestial object. Regardless, comparing the AtLAST sensitivity calculator with what \texttt{maria} outputs is a useful cross-check on results from \texttt{maria}.
            
            The lower panel in Figure~\ref{fig:AtLAST_noise_maps} displays the noise map for our simulated AtLAST observation, but with additional Fourier filtering and a common mode subtraction applied. Within the central $15\arcmin$ region, this map has a beam sensitivity of 12.0 $\mu$Jy/Beam when assuming a diffraction-limited beam shape. In contrast to the AtLAST sensitivity estimator, which provides a sensitivity of $13.97$ $\mu$Jy/Beam for an 8.6~minute long observation scan\footnote{We used the average elevation of $57^\circ$, dual polarization mode, and a H$_2$O profile percentile of 25.}, we find a slightly better per beam sensitivity (i.e., lower RMS) than theoretical predictions even without using the full AtLAST design specifics, and even though the 8.6-minute long integration time is distributed over the map, since the detector array moves on and off source as, for instance, it was for MUSTANG-2 (see Fig~\ref{fig:setup}). 
            Of course, the overall performance of \texttt{maria} depends somewhat on the actual instrument performance, and here we have conservatively adopted the same white and pink noise levels demonstrated by ACT. Regardless, the similar point source sensitivities appear to indicate that the two methodologies broadly agree. 

    \subsection{Synthetic Imaging}\label{sec:forecast_atlast}

        Without delving into the details of cluster astrophysics and the scientific advancements a telescope like AtLAST could bring to the field, we briefly demonstrate how to use \texttt{maria} to make forecasts for upcoming experiments like AtLAST and compare the resulting transfer function with that of MUSTANG-2 (see Section~\ref{sec:M2forcast}).

        Using a similar approach as employed for the MUSTANG-2 case, Figure~\ref{fig:AtLAST_mock-obs} shows the input map, the filtered map, and a synthetic observation of an output from the Dianoga hydrodynamical simulation (see also section~\ref{sec:M2forcast}). The input maps represent the intracluster medium observed via the SZ effect from a cluster of galaxies. As an input, we used two clusters at $z = 0.0688$, with one exhibiting a more compact surface brightness distribution than the other, but both showing features due to cluster merger activities. The chosen redshift corresponds to a $1^\circ$ FoV which corresponds to approximately $4.6$~Mpc, roughly equivalent to $R_{200,c}$. The two clusters, with halo masses of $M_{\rm 500,c} = (8.8,\ 7.8)\times 10^{14}~{\rm M_\odot}$, are shown in that specific order in the first column of Figure~\ref{fig:AtLAST_mock-obs}. 
        
        Because of the low surface brightness and large projection on the sky, the only telescopes currently capable of detecting these clusters in the mm-wave regime are CMB-survey experiments, which typically have resolutions of $\approx1-10\arcmin$ at 100~GHz. Thus, AtLAST-like observations with a resolution of $16\arcsec$ at 92~GHz could open new parameter spaces in cluster astrophysics. Such a telescope will be able to detect and resolve these objects as shown in the middle column of Figure~\ref{fig:AtLAST_mock-obs}.
        These filtered maps indicate that after applying the transfer function, the detailed structures are distinctly visible. Here, we lowered the frequency of the highpass filter in comparison to what we used to make the noise map in Figure~\ref{fig:AtLAST_noise_maps}, to not filter out the flux present at large scales. However, this led to a slight increase of predominantly atmospheric noise in the synthetic observation, as visible in the third and final column. These maps are made from 12 corrupted concatenated scans, each lasting 7.5 minutes, adding to 90 minutes of total integration time per cluster for the assumed instrument, which we emphasize is limited in detector count compared to a future fully populated AtLAST focal plane (see Sect.~\ref{sec:AtLAST_setup}). For both clusters, it is evident that the images are noisier than the map shown in Figure~\ref{fig:AtLAST_noise_maps} because it is more difficult to separate the atmosphere from the extended signal at these scales ($>15\arcmin$). However, both mock observations show clear detections of the interesting resolved features. This highlights the capabilities of a 50-meter single dish telescope like AtLAST with a $0.25^\circ$ FoV equipped with 3000 detectors and a scanning velocity of $0.5^\circ \rm s^{-1}$. 

        \begin{figure}[t]
            \centering
            \includegraphics[width = \hsize]{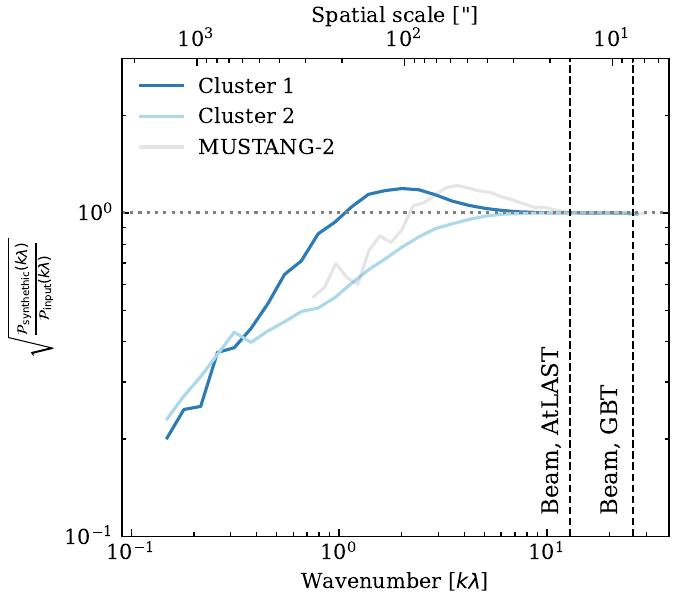}
            \caption{The resulting transfer functions of both clusters with cluster 1 corresponding to the first row in Figure~\ref{fig:AtLAST_mock-obs}. It is derived from the power spectra of the filtered map normalized by that of the input map. In gray, we show the equivalent function but from a MUSTANG-2 observation (see Figure~\ref{fig:tranfer_M2}).}
            \label{fig:Atlast_tranferfunction}
        \end{figure}
        
        The contrast between the transfer function of these observations, shown in Figure~\ref{fig:Atlast_tranferfunction}, with respect to that of MUSTANG-2, also indicates how much better the larger FoV and higher mapping speed of AtLAST are at constraining larger spatial scales. The difference in transfer functions between both clusters comes from the difference in morphology and dynamical state.

\section{Summary and Future Outlook}\label{sec:conclusion}

    \subsection{Summary}

        In this work, we utilized and further developed the atmospheric modeling tools that were first presented in \citet{morris2022} to make forecasts for current and future single-dish facilities operating in the (sub-)mm-wave regime through bolometer arrays. We presented \texttt{maria}, now an open-source telescope simulator, and used it to generate synthetic time ordered data that recreate real observations. Furthermore, we built in a simplistic, Python-based mapmaker to facilitate interpretation for the broad user community. 
        
        The main drivers for this work were threefold. The first motivation is that the broader (sub-)mm community requires an easy-to-use, computationally efficient, and versatile simulator applicable to a variety of science cases, telescope sites, and designs. Section~\ref{sec:method} provided the overview on how \texttt{maria} is set up to fulfill this goal. Second, \texttt{maria} enables a fair comparison between (sub-)mm observations with hydrodynamical simulations and other input models, as illustrated in Section~\ref{sec:valid}. Henceforth, we can validate observational tools with these simulations. In particular, since the field of resolved SZ studies is now becoming more established, thanks to, for example, the wider bandwidth and sensitivity upgrades to ALMA \citep{Carpenter2023} and the increased sensitivity of large single dish facilities, comparisons with hydrodynamical simulations are necessary to understand the systematic uncertainties in SZ analyses. The third and final motivation was to establish a connection between telescope designs, such as the number of detectors, field of view, and mapping speed, and the scientific goals for future facilities like the 50~meter Atacama Large Aperture Submillimeter Telescope (AtLAST). This objective was addressed in Section~\ref{sec:AtLAST}. However, regarding the latter goal, we simplified the instrument's capabilities to facilitate simulation. More comprehensive, large-scale forecasts will be presented in follow-up works, utilizing the AtLAST design's full field of view, polarization, and multichroic capabilities. For comparison, the data rates from mock observations will be $\approx 64\times$ larger when simulating a 2$^\circ$ FoV monochromatic instrument when simply filling AtLAST's focal plane. This highlights the fact that AtLAST, with instrumentation fully populating the field of view, will be firmly situated in the Big Data regime.     
    
        As this work illustrates in general the success of \texttt{maria}, it also has its limitations. Cosmic Microwave Background experiments or other single-dish simulators used for, for instance, the TolTEC Camera \citep[see, e.g.,][]{Bryan2018} often refer to the Time-Ordered Astrophysics Scalable Tools package \citep[\texttt{TOAST},][]{Puglisi2021,Kisner2023} or \texttt{BoloCalc} \citep{Hill2018} to study systematic errors induced by the beam and calibration uncertainties. 
        Apart from correlated atmospheric noise, \texttt{TOAST} and \texttt{BoloCalc} can also simulate cross-talk between detectors and scan-synchronous signals due to ground pickup as well as simulate leakage from total intensity to polarized light \citep{Hill2018, Dachlythra2024}. These noise contributions are not implemented in \texttt{maria}, nor are, for instance, surface deformations of the dish caused by gravity and thermal fluctuations. With \texttt{maria}, we simplistically assume diffraction-limited beams and uncorrelated pink and white noise. However, as demonstrated in Section~\ref{sec:valid}, the noise corruption model implemented effectively replicates the noise characteristics observed in MUSTANG-2 data. Moreover, it is easily adaptable to other telescope configurations at different sites. Additionally, our implementation of the location-and-time-specific atmosphere in \texttt{maria} stands out for its uniqueness while maintaining computational efficiency in computing the full evolving atmospheric volume.

    \subsection{Future outlook}

        In the forthcoming era of big data and the global shift towards more sustainable astronomy, there is an urgent need to develop green telescopes that prioritize energy efficiency in both hardware and software. As mentioned, facilities like AtLAST, along with SO \citep{Ade2019}, SPT-3G \citep{Benson2014}, CMB-S4 \citep{CMBS42016}, SKA \citep{Dewdney2009}, and ngVLA \citep{Murphy2018} will generate large data volumes and require substantial storage and supercomputing resources to process and calibrate the data. One straightforward method to reduce computation time and storage requirements is to discard data, such as visibilities, or to down-sample time streams (as already done with MUSTANG-2 observations). However, these approaches could impact the reproducibility of research and limit the effectiveness of tools designed to extract the maximum scientific value from datasets.
        The above considerations underscore the need for more efficient reconstruction methods. In the case of \texttt{maria}, we will make significant improvements to the code infrastructure, such as enhancing parallelization and GPU acceleration by utilizing \texttt{dask} \citep{rocklin2015dask} and \texttt{jax} \citep{jax2018github}, which can boost performance by one or two orders of magnitude. This should make \texttt{maria} computationally efficient enough to simulate AtLAST and upcoming CMB-survey facilities to their full capacities. 
        
        In general, we expect that the most significant software advancement in the processing of ground-based CMB data will be achieved through the use of Gaussian processes. Although machine learning implementations might have a higher energy cost than simpler analysis tools, such as the naive mapmaking used in this work, they might be more efficient for retrieving scientific information and distinguishing the astronomical source from noise components. Single-dish facilities face significant challenges in separating atmospheric and astronomical signals, especially at large scales. This difficulty arises due to the large dynamic range in amplitudes—approximately seven orders of magnitude—between the atmosphere and the astronomical signal, as well as the fact that they share common modes across detectors, thus plausibly confusing large-scale signals for a common mode measured in the atmosphere. Additionally, the nonlinearity caused by sky scanning with a non-constant velocity further complicates the transformation from time-domain data to map-domain data. Traditional mapmaking methods assume a stationary, flat sky and pointing matrix creating a linear, invertible operation from time streams to sky coordinates \citep[see,][section 11 for a more thorough description of mapmaking]{Dunner2013}. However, this assumption is not valid at, for instance, the turnaround of a back-and-forth scan, eventually leading to loss of signal on large scales \citep{Naess2023}. Advancing map-making techniques and the corresponding steps for denoising timestreams from millions of detectors are well-suited to applications in information field theory, such as those relying on the Numerical Information Field Theory tool (\texttt{NIFTy}; \citealt{Selig2014}, \citealt{Torsten2019}, \citealt{Edenhofer2024J}). \texttt{NIFTy} provides a powerful framework for inferring the true signal, which is obscured by a response function. In this context, it involves sampling the true signal and convolving the resulting detector time series with contributions from the atmosphere, the astronomical signal, and a noise component. The component separation in \texttt{maria} enables \texttt{NIFTy} to learn how to separate different components of the time series while accounting for non-linearity in sky sampling and the evolving atmosphere. These developments are currently in progress and will be explored and discussed in future work (W\"{u}rzinger et al., in prep.).
        
        A final application of \texttt{maria} to discuss is its potential for studying phase delays caused by the ever-changing turbulent atmosphere. This is particularly relevant for long-baseline, high-frequency campaigns conducted with ALMA, as \texttt{maria} could serve as an improved phase screen model, adding realism to replicate the empirical findings on measured phase stabilities reported in \citet{Maud2023}. These findings are significant not only for ALMA but also for the ngVLA and would enable proper comparison of the \texttt{maria} code and observations in the terahertz regime, something currently unattainable with most single-dish facilities. Future implementations of \texttt{maria} will also be better equipped to simulate direct detection spectrometers as was done, for instance, in \texttt{TiEMPO} \citep{Huijten2022}. Currently, we advise creating ``moment-0'' maps with a small frequency width, while future versions will be able to generate mock observations to imitate observations using heterodyne and integrated field units with higher spectral resolution.

        In closing, we emphasize that this paper is meant to introduce a virtual (sub-)mm telescope simulation tool that is generic and applicable across several science fields, from observations of the Sun (Kirkaune et al., in prep) to the circumgalactic medium of galaxies \citep[see, e.g.,][]{Lee2024, Schimek+24}. Even though the synthetic observations in this work all show the intracluster medium as seen via the SZ effect, the tool is broadly applicable and will be featured in future work.
 
\section*{Acknowledgements}
    This project has received funding from the European Union’s Horizon 2020 research and innovation program under grant agreement No.\ 951815 (AtLAST).
    MUSTANG-2 is supported by the NSF award number 1615604, by the Mt.\ Cuba Astronomical Foundation, and the Green Bank Observatory. 
    The Green Bank Observatory is a facility of the National Science Foundation operated under cooperative agreement by Associated Universities, Inc.
    T.\ Morris acknowledges the support of L.\ Page.
    T.\ Morris acknowledges the musical work of Alan J.\ Lerner and Frederick Loewe as inspiration for the naming of the \texttt{maria} code.
    This work has been supported by the French government, through the UCA\textsuperscript{J.E.D.I.} Investments in the Future project managed by the National Research Agency (ANR) with the reference number ANR-15-IDEX-01. L.D.M.\ is supported by the ERC-StG ``ClustersXCosmo'' grant agreement 716762. L.D.M.\ further acknowledges financial contribution from the agreement ASI-INAF n.2017-14-H.0. 
    JW is funded by the Deutsche Forschungsgemeinschaft (DFG, German Research Foundation) under Germany's Excellence Strategy – EXC 2094 – 390783311.

\bibliographystyle{maria} 
\bibliography{maria} 

\begin{thebibliography}{81}
\expandafter\ifx\csname natexlab\endcsname\relax\def\natexlab#1{#1}\fi

\bibitem[{{Abazajian} {et~al.}(2016){Abazajian}, {Adshead}, {Ahmed}, {Allen},
  {Alonso}, {Arnold}, {Baccigalupi}, {Bartlett}, {Battaglia}, {Benson},
  {Bischoff}, {Borrill}, {Buza}, {Calabrese}, {Caldwell}, {Carlstrom}, {Chang},
  {Crawford}, {Cyr-Racine}, {De Bernardis}, {de Haan}, {di Serego Alighieri},
  {Dunkley}, {Dvorkin}, {Errard}, {Fabbian}, {Feeney}, {Ferraro}, {Filippini},
  {Flauger}, {Fuller}, {Gluscevic}, {Green}, {Grin}, {Grohs}, {Henning},
  {Hill}, {Hlozek}, {Holder}, {Holzapfel}, {Hu}, {Huffenberger}, {Keskitalo},
  {Knox}, {Kosowsky}, {Kovac}, {Kovetz}, {Kuo}, {Kusaka}, {Le Jeune}, {Lee},
  {Lilley}, {Loverde}, {Madhavacheril}, {Mantz}, {Marsh}, {McMahon},
  {Meerburg}, {Meyers}, {Miller}, {Munoz}, {Nguyen}, {Niemack}, {Peloso},
  {Peloton}, {Pogosian}, {Pryke}, {Raveri}, {Reichardt}, {Rocha}, {Rotti},
  {Schaan}, {Schmittfull}, {Scott}, {Sehgal}, {Shandera}, {Sherwin}, {Smith},
  {Sorbo}, {Starkman}, {Story}, {van Engelen}, {Vieira}, {Watson}, {Whitehorn},
  \& {Kimmy Wu}}]{CMBS42016}
{Abazajian}, K.~N., {Adshead}, P., {Ahmed}, Z., {et~al.} 2016, arXiv e-prints,
  arXiv:1610.02743

\bibitem[{{Ade} {et~al.}(2019){Ade}, {Aguirre}, {Ahmed}, {Aiola}, {Ali},
  {Alonso}, {Alvarez}, {Arnold}, {Ashton}, {Austermann}, {Awan}, {Baccigalupi},
  {Baildon}, {Barron}, {Battaglia}, {Battye}, {Baxter}, {Bazarko}, {Beall},
  {Bean}, {Beck}, {Beckman}, {Beringue}, {Bianchini}, {Boada}, {Boettger},
  {Bond}, {Borrill}, {Brown}, {Bruno}, {Bryan}, {Calabrese}, {Calafut},
  {Calisse}, {Carron}, {Challinor}, {Chesmore}, {Chinone}, {Chluba}, {Cho},
  {Choi}, {Coppi}, {Cothard}, {Coughlin}, {Crichton}, {Crowley}, {Crowley},
  {Cukierman}, {D'Ewart}, {D{\"u}nner}, {de Haan}, {Devlin}, {Dicker},
  {Didier}, {Dobbs}, {Dober}, {Duell}, {Duff}, {Duivenvoorden}, {Dunkley},
  {Dusatko}, {Errard}, {Fabbian}, {Feeney}, {Ferraro}, {Flux{\`a}}, {Freese},
  {Frisch}, {Frolov}, {Fuller}, {Fuzia}, {Galitzki}, {Gallardo}, {Tomas Galvez
  Ghersi}, {Gao}, {Gawiser}, {Gerbino}, {Gluscevic}, {Goeckner-Wald}, {Golec},
  {Gordon}, {Gralla}, {Green}, {Grigorian}, {Groh}, {Groppi}, {Guan},
  {Gudmundsson}, {Han}, {Hargrave}, {Hasegawa}, {Hasselfield}, {Hattori},
  {Haynes}, {Hazumi}, {He}, {Healy}, {Henderson}, {Hervias-Caimapo}, {Hill},
  {Hill}, {Hilton}, {Hilton}, {Hincks}, {Hinshaw}, {Hlo{\v{z}}ek}, {Ho}, {Ho},
  {Howe}, {Huang}, {Hubmayr}, {Huffenberger}, {Hughes}, {Ijjas}, {Ikape},
  {Irwin}, {Jaffe}, {Jain}, {Jeong}, {Kaneko}, {Karpel}, {Katayama}, {Keating},
  {Kernasovskiy}, {Keskitalo}, {Kisner}, {Kiuchi}, {Klein}, {Knowles},
  {Koopman}, {Kosowsky}, {Krachmalnicoff}, {Kuenstner}, {Kuo}, {Kusaka},
  {Lashner}, {Lee}, {Lee}, {Leon}, {Leung}, {Lewis}, {Li}, {Li}, {Limon},
  {Linder}, {Lopez-Caraballo}, {Louis}, {Lowry}, {Lungu}, {Madhavacheril},
  {Mak}, {Maldonado}, {Mani}, {Mates}, {Matsuda}, {Maurin}, {Mauskopf}, {May},
  {McCallum}, {McKenney}, {McMahon}, {Meerburg}, {Meyers}, {Miller},
  {Mirmelstein}, {Moodley}, {Munchmeyer}, {Munson}, {Naess}, {Nati},
  {Navaroli}, {Newburgh}, {Nguyen}, {Niemack}, {Nishino}, {Orlowski-Scherer},
  {Page}, {Partridge}, {Peloton}, {Perrotta}, {Piccirillo}, {Pisano},
  {Poletti}, {Puddu}, {Puglisi}, {Raum}, {Reichardt}, {Remazeilles},
  {Rephaeli}, {Riechers}, {Rojas}, {Roy}, {Sadeh}, {Sakurai}, {Salatino},
  {Sathyanarayana Rao}, {Schaan}, {Schmittfull}, {Sehgal}, {Seibert}, {Seljak},
  {Sherwin}, {Shimon}, {Sierra}, {Sievers}, {Sikhosana}, {Silva-Feaver},
  {Simon}, {Sinclair}, {Siritanasak}, {Smith}, {Smith}, {Spergel}, {Staggs},
  {Stein}, {Stevens}, {Stompor}, {Suzuki}, {Tajima}, {Takakura}, {Teply},
  {Thomas}, {Thorne}, {Thornton}, {Trac}, {Tsai}, {Tucker}, {Ullom},
  {Vagnozzi}, {van Engelen}, {Van Lanen}, {Van Winkle}, {Vavagiakis},
  {Verg{\`e}s}, {Vissers}, {Wagoner}, {Walker}, {Ward}, {Westbrook},
  {Whitehorn}, {Williams}, {Williams}, {Wollack}, {Xu}, {Yu}, {Yu}, {Zago},
  {Zhang}, {Zhu}, \& {Simons Observatory Collaboration}}]{Ade2019}
{Ade}, P., {Aguirre}, J., {Ahmed}, Z., {et~al.} 2019, \jcap, 2019, 056

\bibitem[{Ass{\'e}mat {et~al.}(2006)Ass{\'e}mat, Wilson, \&
  Gendron}]{assemat2006method}
Ass{\'e}mat, F., Wilson, R.~W., \& Gendron, E. 2006, Optics express, 14, 988

\bibitem[{{Benson} {et~al.}(2014){Benson}, {Ade}, {Ahmed}, {Allen}, {Arnold},
  {Austermann}, {Bender}, {Bleem}, {Carlstrom}, {Chang}, {Cho}, {Cliche},
  {Crawford}, {Cukierman}, {de Haan}, {Dobbs}, {Dutcher}, {Everett}, {Gilbert},
  {Halverson}, {Hanson}, {Harrington}, {Hattori}, {Henning}, {Hilton},
  {Holder}, {Holzapfel}, {Irwin}, {Keisler}, {Knox}, {Kubik}, {Kuo}, {Lee},
  {Leitch}, {Li}, {McDonald}, {Meyer}, {Montgomery}, {Myers}, {Natoli},
  {Nguyen}, {Novosad}, {Padin}, {Pan}, {Pearson}, {Reichardt}, {Ruhl},
  {Saliwanchik}, {Simard}, {Smecher}, {Sayre}, {Shirokoff}, {Stark}, {Story},
  {Suzuki}, {Thompson}, {Tucker}, {Vanderlinde}, {Vieira}, {Vikhlinin}, {Wang},
  {Yefremenko}, \& {Yoon}}]{Benson2014}
{Benson}, B.~A., {Ade}, P.~A.~R., {Ahmed}, Z., {et~al.} 2014, in Society of
  Photo-Optical Instrumentation Engineers (SPIE) Conference Series, Vol. 9153,
  Millimeter, Submillimeter, and Far-Infrared Detectors and Instrumentation for
  Astronomy VII, ed. W.~S. {Holland} \& J.~{Zmuidzinas}, 91531P

\bibitem[{{Biffi} {et~al.}(2011){Biffi}, {Dolag}, {Boehringer}, \&
  {Lemson}}]{Biffi2011}
{Biffi}, V., {Dolag}, K., {Boehringer}, H., \& {Lemson}, G. 2011, {PHOX: X-ray
  Photon Simulator}, Astrophysics Source Code Library, record ascl:1112.004

\bibitem[{{Bleem} {et~al.}(2020){Bleem}, {Bocquet}, {Stalder}, {Gladders},
  {Ade}, {Allen}, {Anderson}, {Annis}, {Ashby}, {Austermann}, {Avila}, {Avva},
  {Bayliss}, {Beall}, {Bechtol}, {Bender}, {Benson}, {Bertin}, {Bianchini},
  {Blake}, {Brodwin}, {Brooks}, {Buckley-Geer}, {Burke}, {Carlstrom}, {Rosell},
  {Carrasco Kind}, {Carretero}, {Chang}, {Chiang}, {Citron}, {Moran},
  {Costanzi}, {Crawford}, {Crites}, {da Costa}, {de Haan}, {De Vicente},
  {Desai}, {Diehl}, {Dietrich}, {Dobbs}, {Eifler}, {Everett}, {Flaugher},
  {Floyd}, {Frieman}, {Gallicchio}, {Garc{\'\i}a-Bellido}, {George}, {Gerdes},
  {Gilbert}, {Gruen}, {Gruendl}, {Gschwend}, {Gupta}, {Gutierrez}, {Halverson},
  {Harrington}, {Henning}, {Heymans}, {Holder}, {Hollowood}, {Holzapfel},
  {Honscheid}, {Hrubes}, {Huang}, {Hubmayr}, {Irwin}, {James}, {Jeltema},
  {Joudaki}, {Khullar}, {Klein}, {Knox}, {Kuropatkin}, {Lee}, {Li}, {Lidman},
  {Lowitz}, {MacCrann}, {Mahler}, {Maia}, {Marshall}, {McDonald}, {McMahon},
  {Melchior}, {Menanteau}, {Meyer}, {Miquel}, {Mocanu}, {Mohr}, {Montgomery},
  {Nadolski}, {Natoli}, {Nibarger}, {Noble}, {Novosad}, {Padin}, {Palmese},
  {Parkinson}, {Patil}, {Paz-Chinch{\'o}n}, {Plazas}, {Pryke}, {Ramachandra},
  {Reichardt}, {Remolina Gonz{\'a}lez}, {Romer}, {Roodman}, {Ruhl}, {Rykoff},
  {Saliwanchik}, {Sanchez}, {Saro}, {Sayre}, {Schaffer}, {Schrabback},
  {Serrano}, {Sharon}, {Sievers}, {Smecher}, {Smith}, {Soares-Santos}, {Stark},
  {Story}, {Suchyta}, {Tarle}, {Tucker}, {Vanderlinde}, {Veach}, {Vieira},
  {Wang}, {Weller}, {Whitehorn}, {Wu}, {Yefremenko}, \& {Zhang}}]{Bleem2020}
{Bleem}, L.~E., {Bocquet}, S., {Stalder}, B., {et~al.} 2020, \apjs, 247, 25

\bibitem[{{Bleem} {et~al.}(2015){Bleem}, {Stalder}, {de Haan}, {Aird}, {Allen},
  {Applegate}, {Ashby}, {Bautz}, {Bayliss}, {Benson}, {Bocquet}, {Brodwin},
  {Carlstrom}, {Chang}, {Chiu}, {Cho}, {Clocchiatti}, {Crawford}, {Crites},
  {Desai}, {Dietrich}, {Dobbs}, {Foley}, {Forman}, {George}, {Gladders},
  {Gonzalez}, {Halverson}, {Hennig}, {Hoekstra}, {Holder}, {Holzapfel},
  {Hrubes}, {Jones}, {Keisler}, {Knox}, {Lee}, {Leitch}, {Liu}, {Lueker},
  {Luong-Van}, {Mantz}, {Marrone}, {McDonald}, {McMahon}, {Meyer}, {Mocanu},
  {Mohr}, {Murray}, {Padin}, {Pryke}, {Reichardt}, {Rest}, {Ruel}, {Ruhl},
  {Saliwanchik}, {Saro}, {Sayre}, {Schaffer}, {Schrabback}, {Shirokoff},
  {Song}, {Spieler}, {Stanford}, {Staniszewski}, {Stark}, {Story}, {Stubbs},
  {Vanderlinde}, {Vieira}, {Vikhlinin}, {Williamson}, {Zahn}, \&
  {Zenteno}}]{Bleem2015}
{Bleem}, L.~E., {Stalder}, B., {de Haan}, T., {et~al.} 2015, \apjs, 216, 27

\bibitem[{{Bonanomi} {et~al.}(2024){Bonanomi}, {Hacar}, {Socci}, {Petry}, \&
  {Suri}}]{Bonanomi2024}
{Bonanomi}, F., {Hacar}, A., {Socci}, A., {Petry}, D., \& {Suri}, S. 2024,
  \aap, 688, A30

\bibitem[{{Booth} {et~al.}(2024){Booth}, {Klaassen}, {Cicone}, {Mroczkowski},
  {Wedemeyer}, {Akiyama}, {Bower}, {Cordiner}, {Di Mascolo}, {Johnstone}, {van
  Kampen}, {Lee}, {Liu}, {Orlowski-Scherer}, {Saintonge}, {Smith}, \&
  {Thelen}}]{Booth2024}
{Booth}, M., {Klaassen}, P., {Cicone}, C., {et~al.} 2024, in Society of
  Photo-Optical Instrumentation Engineers (SPIE) Conference Series, Vol. 13102,
  Society of Photo-Optical Instrumentation Engineers (SPIE) Conference Series,
  ed. J.~{Zmuidzinas} \& J.-R. {Gao}, 1310206

\bibitem[{Bradbury {et~al.}(2018)Bradbury, Frostig, Hawkins, Johnson, Leary,
  Maclaurin, Necula, Paszke, Vander{P}las, Wanderman-{M}ilne, \&
  Zhang}]{jax2018github}
Bradbury, J., Frostig, R., Hawkins, P., {et~al.} 2018, {JAX}: composable
  transformations of {P}ython+{N}um{P}y programs

\bibitem[{{Bryan} {et~al.}(2018){Bryan}, {Austermann}, {Ferrusca}, {Mauskopf},
  {McMahon}, {Monta{\~n}a}, {Simon}, {Novak}, {S{\'a}nchez-Arg{\"u}elles}, \&
  {Wilson}}]{Bryan2018}
{Bryan}, S., {Austermann}, J., {Ferrusca}, D., {et~al.} 2018, in Society of
  Photo-Optical Instrumentation Engineers (SPIE) Conference Series, Vol. 10708,
  Millimeter, Submillimeter, and Far-Infrared Detectors and Instrumentation for
  Astronomy IX, ed. J.~{Zmuidzinas} \& J.-R. {Gao}, 107080J

\bibitem[{{Carlstrom} {et~al.}(2011){Carlstrom}, {Ade}, {Aird}, {Benson},
  {Bleem}, {Busetti}, {Chang}, {Chauvin}, {Cho}, {Crawford}, {Crites}, {Dobbs},
  {Halverson}, {Heimsath}, {Holzapfel}, {Hrubes}, {Joy}, {Keisler}, {Lanting},
  {Lee}, {Leitch}, {Leong}, {Lu}, {Lueker}, {Luong-Van}, {McMahon}, {Mehl},
  {Meyer}, {Mohr}, {Montroy}, {Padin}, {Plagge}, {Pryke}, {Ruhl}, {Schaffer},
  {Schwan}, {Shirokoff}, {Spieler}, {Staniszewski}, {Stark}, {Tucker},
  {Vanderlinde}, {Vieira}, \& {Williamson}}]{Carlstrom2011}
{Carlstrom}, J.~E., {Ade}, P.~A.~R., {Aird}, K.~A., {et~al.} 2011, \pasp, 123,
  568

\bibitem[{{Carpenter} {et~al.}(2023){Carpenter}, {Brogan}, {Iono}, \&
  {Mroczkowski}}]{Carpenter2023}
{Carpenter}, J., {Brogan}, C., {Iono}, D., \& {Mroczkowski}, T. 2023, in
  Physics and Chemistry of Star Formation: The Dynamical ISM Across Time and
  Spatial Scales, 304

\bibitem[{{CASA Team} {et~al.}(2022){CASA Team}, {Bean}, {Bhatnagar}, {Castro},
  {Donovan Meyer}, {Emonts}, {Garcia}, {Garwood}, {Golap}, {Gonzalez Villalba},
  {Harris}, {Hayashi}, {Hoskins}, {Hsieh}, {Jagannathan}, {Kawasaki},
  {Keimpema}, {Kettenis}, {Lopez}, {Marvil}, {Masters}, {McNichols},
  {Mehringer}, {Miel}, {Moellenbrock}, {Montesino}, {Nakazato}, {Ott}, {Petry},
  {Pokorny}, {Raba}, {Rau}, {Schiebel}, {Schweighart}, {Sekhar}, {Shimada},
  {Small}, {Steeb}, {Sugimoto}, {Suoranta}, {Tsutsumi}, {van Bemmel},
  {Verkouter}, {Wells}, {Xiong}, {Szomoru}, {Griffith}, {Glendenning}, \&
  {Kern}}]{CASA2022}
{CASA Team}, {Bean}, B., {Bhatnagar}, S., {et~al.} 2022, \pasp, 134, 114501

\bibitem[{{CCAT-Prime Collaboration}(2023)}]{CCAT2023}
{CCAT-Prime Collaboration}. 2023, \apjs, 264, 7

\bibitem[{Conan {et~al.}(2000)Conan, Borgnino, Ziad, \&
  Martin}]{conan2000analytical}
Conan, R., Borgnino, J., Ziad, A., \& Martin, F. 2000, JOSA A, 17, 1807

\bibitem[{Consortini \& Ronchi(1972)}]{consortini1972choice}
Consortini, A. \& Ronchi, L. 1972, Applied Optics, 11, 1205

\bibitem[{{Coulton} {et~al.}(2024){Coulton}, {Madhavacheril}, {Duivenvoorden},
  {Hill}, {Abril-Cabezas}, {Ade}, {Aiola}, {Alford}, {Amiri}, {Amodeo}, {An},
  {Atkins}, {Austermann}, {Battaglia}, {Battistelli}, {Beall}, {Bean},
  {Beringue}, {Bhandarkar}, {Biermann}, {Bolliet}, {Bond}, {Cai}, {Calabrese},
  {Calafut}, {Capalbo}, {Carrero}, {Chesmore}, {Cho}, {Choi}, {Clark},
  {Rosado}, {Cothard}, {Coughlin}, {Crowley}, {Devlin}, {Dicker}, {Doze},
  {Duell}, {Duff}, {Dunkley}, {D{\"u}nner}, {Fanfani}, {Fankhanel}, {Farren},
  {Ferraro}, {Freundt}, {Fuzia}, {Gallardo}, {Garrido}, {Givans}, {Gluscevic},
  {Golec}, {Guan}, {Halpern}, {Han}, {Hasselfield}, {Healy}, {Henderson},
  {Hensley}, {Herv{\'\i}as-Caimapo}, {Hilton}, {Hilton}, {Hincks},
  {Hlo{\v{z}}ek}, {Ho}, {Huber}, {Hubmayr}, {Huffenberger}, {Hughes}, {Irwin},
  {Isopi}, {Jense}, {Keller}, {Kim}, {Knowles}, {Koopman}, {Kosowsky},
  {Kramer}, {Kusiak}, {La Posta}, {Lakey}, {Lee}, {Li}, {Li}, {Limon},
  {Lokken}, {Louis}, {Lungu}, {MacCrann}, {MacInnis}, {Maldonado}, {Maldonado},
  {Mallaby-Kay}, {Marques}, {van Marrewijk}, {McCarthy}, {McMahon}, {Mehta},
  {Menanteau}, {Moodley}, {Morris}, {Mroczkowski}, {Naess}, {Namikawa}, {Nati},
  {Newburgh}, {Nicola}, {Niemack}, {Nolta}, {Orlowski-Scherer}, {Page},
  {Pandey}, {Partridge}, {Prince}, {Puddu}, {Qu}, {Radiconi}, {Robertson},
  {Rojas}, {Sakuma}, {Salatino}, {Schaan}, {Schmitt}, {Sehgal}, {Shaikh},
  {Sherwin}, {Sierra}, {Sievers}, {Sif{\'o}n}, {Simon}, {Sonka}, {Spergel},
  {Staggs}, {Storer}, {Switzer}, {Tampier}, {Thornton}, {Trac}, {Treu},
  {Tucker}, {Ullom}, {Vale}, {Van Engelen}, {Van Lanen}, {Vargas},
  {Vavagiakis}, {Wagoner}, {Wang}, {Wenzl}, {Wollack}, {Xu}, {Zago}, \&
  {Zheng}}]{Coulton2024}
{Coulton}, W., {Madhavacheril}, M.~S., {Duivenvoorden}, A.~J., {et~al.} 2024,
  \prd, 109, 063530

\bibitem[{{Dachlythra} {et~al.}(2024){Dachlythra}, {Duivenvoorden},
  {Gudmundsson}, {Hasselfield}, {Coppi}, {Adler}, {Alonso}, {Azzoni},
  {Chesmore}, {Fabbian}, {Ganga}, {Gerras}, {Jaffe}, {Johnson}, {Keating},
  {Keskitalo}, {Kisner}, {Krachmalnicoff}, {Lungu}, {Matsuda}, {Naess}, {Page},
  {Puddu}, {Puglisi}, {Simon}, {Teply}, {Tsan}, {Wollack}, {Wolz}, \&
  {Xu}}]{Dachlythra2024}
{Dachlythra}, N., {Duivenvoorden}, A.~J., {Gudmundsson}, J.~E., {et~al.} 2024,
  \apj, 961, 138

\bibitem[{{Devlin} {et~al.}(2009){Devlin}, {Ade}, {Aretxaga}, {Bock}, {Chapin},
  {Griffin}, {Gundersen}, {Halpern}, {Hargrave}, {Hughes}, {Klein}, {Marsden},
  {Martin}, {Mauskopf}, {Moncelsi}, {Netterfield}, {Ngo}, {Olmi}, {Pascale},
  {Patanchon}, {Rex}, {Scott}, {Semisch}, {Thomas}, {Truch}, {Tucker},
  {Tucker}, {Viero}, \& {Wiebe}}]{Devlin2009}
{Devlin}, M.~J., {Ade}, P. A.~R., {Aretxaga}, I., {et~al.} 2009, \nat, 458, 737

\bibitem[{{Dewdney} {et~al.}(2009){Dewdney}, {Hall}, {Schilizzi}, \&
  {Lazio}}]{Dewdney2009}
{Dewdney}, P.~E., {Hall}, P.~J., {Schilizzi}, R.~T., \& {Lazio}, T.~J.~L.~W.
  2009, IEEE Proceedings, 97, 1482

\bibitem[{{Di Mascolo} {et~al.}(2024){Di Mascolo}, {Perrott}, {Mroczkowski},
  {Andreon}, {Ettori}, {Simionescu}, {Raghunathan}, {van Marrewijk}, {Cicone},
  {Lee}, {Nelson}, {Sommovigo}, {Booth}, {Klaassen}, {Andreani}, {Cordiner},
  {Johnstone}, {van Kampen}, {Liu}, {Maccarone}, {Morris}, {Saintonge},
  {Smith}, {Thelen}, \& {Wedemeyer}}]{DiMascolo2024}
{Di Mascolo}, L., {Perrott}, Y., {Mroczkowski}, T., {et~al.} 2024, arXiv
  e-prints, arXiv:2403.00909

\bibitem[{{Di Mascolo} {et~al.}(2023){Di Mascolo}, {Saro}, {Mroczkowski},
  {Borgani}, {Churazov}, {Rasia}, {Tozzi}, {Dannerbauer}, {Basu}, {Carilli},
  {Ginolfi}, {Miley}, {Nonino}, {Pannella}, {Pentericci}, \&
  {Rizzo}}]{DiMascolo2023}
{Di Mascolo}, L., {Saro}, A., {Mroczkowski}, T., {et~al.} 2023, \nat, 615, 809

\bibitem[{{Dicker} {et~al.}(2014){Dicker}, {Ade}, {Aguirre}, {Brevik}, {Cho},
  {Datta}, {Devlin}, {Dober}, {Egan}, {Ford}, {Ford}, {Hilton}, {Hubmayr},
  {Irwin}, {Mason}, {Marganian}, {Mello}, {McMahon}, {Mroczkowski}, {Romero},
  {Stanchfield}, {Tucker}, {Vale}, {White}, {Whitehead}, \&
  {Young}}]{Dicker2014}
{Dicker}, S.~R., {Ade}, P.~A.~R., {Aguirre}, J., {et~al.} 2014, in Society of
  Photo-Optical Instrumentation Engineers (SPIE) Conference Series, Vol. 9153,
  Millimeter, Submillimeter, and Far-Infrared Detectors and Instrumentation for
  Astronomy VII, ed. W.~S. {Holland} \& J.~{Zmuidzinas}, 91530J

\bibitem[{{D{\"u}nner} {et~al.}(2013){D{\"u}nner}, {Hasselfield}, {Marriage},
  {Sievers}, {Acquaviva}, {Addison}, {Ade}, {Aguirre}, {Amiri}, {Appel},
  {Barrientos}, {Battistelli}, {Bond}, {Brown}, {Burger}, {Calabrese},
  {Chervenak}, {Das}, {Devlin}, {Dicker}, {Bertrand Doriese}, {Dunkley},
  {Essinger-Hileman}, {Fisher}, {Gralla}, {Fowler}, {Hajian}, {Halpern},
  {Hern{\'a}ndez-Monteagudo}, {Hilton}, {Hilton}, {Hincks}, {Hlozek},
  {Huffenberger}, {Hughes}, {Hughes}, {Infante}, {Irwin}, {Baptiste Juin},
  {Kaul}, {Klein}, {Kosowsky}, {Lau}, {Limon}, {Lin}, {Louis}, {Lupton},
  {Marsden}, {Martocci}, {Mauskopf}, {Menanteau}, {Moodley}, {Moseley},
  {Netterfield}, {Niemack}, {Nolta}, {Page}, {Parker}, {Partridge}, {Quintana},
  {Reid}, {Sehgal}, {Sherwin}, {Spergel}, {Staggs}, {Swetz}, {Switzer},
  {Thornton}, {Trac}, {Tucker}, {Warne}, {Wilson}, {Wollack}, \&
  {Zhao}}]{Dunner2013}
{D{\"u}nner}, R., {Hasselfield}, M., {Marriage}, T.~A., {et~al.} 2013, \apj,
  762, 10

\bibitem[{{Edenhofer} {et~al.}(2024){Edenhofer}, {Frank}, {Roth}, {Leike},
  {Guerdi}, {Scheel-Platz}, {Guardiani}, {Eberle}, {Westerkamp}, \&
  {En{\ss}lin}}]{Edenhofer2024J}
{Edenhofer}, G., {Frank}, P., {Roth}, J., {et~al.} 2024, The Journal of Open
  Source Software, 9, 6593

\bibitem[{{En{\ss}lin}(2019)}]{Torsten2019}
{En{\ss}lin}, T.~A. 2019, Annalen der Physik, 531, 1800127

\bibitem[{{Errard} {et~al.}(2015){Errard}, {Ade}, {Akiba}, {Arnold}, {Atlas},
  {Baccigalupi}, {Barron}, {Boettger}, {Borrill}, {Chapman}, {Chinone},
  {Cukierman}, {Delabrouille}, {Dobbs}, {Ducout}, {Elleflot}, {Fabbian},
  {Feng}, {Feeney}, {Gilbert}, {Goeckner-Wald}, {Halverson}, {Hasegawa},
  {Hattori}, {Hazumi}, {Hill}, {Holzapfel}, {Hori}, {Inoue}, {Jaehnig},
  {Jaffe}, {Jeong}, {Katayama}, {Kaufman}, {Keating}, {Kermish}, {Keskitalo},
  {Kisner}, {Le Jeune}, {Lee}, {Leitch}, {Leon}, {Linder}, {Matsuda},
  {Matsumura}, {Miller}, {Myers}, {Navaroli}, {Nishino}, {Okamura}, {Paar},
  {Peloton}, {Poletti}, {Puglisi}, {Rebeiz}, {Reichardt}, {Richards}, {Ross},
  {Rotermund}, {Schenck}, {Sherwin}, {Siritanasak}, {Smecher}, {Stebor},
  {Steinbach}, {Stompor}, {Suzuki}, {Tajima}, {Takakura}, {Tikhomirov},
  {Tomaru}, {Whitehorn}, {Wilson}, {Yadav}, \& {Zahn}}]{errard2015}
{Errard}, J., {Ade}, P.~A.~R., {Akiba}, Y., {et~al.} 2015, \apj, 809, 63

\bibitem[{{Frayer}(2017)}]{Frayer2017}
{Frayer}, D.~T. 2017, arXiv e-prints, arXiv:1706.02726

\bibitem[{{Frayer} {et~al.}(2019){Frayer}, {Maddalena}, {White}, {Watts},
  {Kepley}, {Li}, \& {Harris}}]{Frayer2019}
{Frayer}, D.~T., {Maddalena}, R.~J., {White}, S., {et~al.} 2019, {Calibration
  of Argus and the 4mm Receiver on the GBT}, Green Bank Telescope Memorandum
  302, June 5, 2019, 21 pages

\bibitem[{{Gallardo} {et~al.}(2024){Gallardo}, {Puddu}, {Mroczkowski}, {Timpe},
  {Dubois-dit-Bonclaude}, {Groh}, {Reichert}, {Cicone}, \&
  {Kaercher}}]{Gallardo2024}
{Gallardo}, P.~A., {Puddu}, R., {Mroczkowski}, T., {et~al.} 2024, arXiv
  e-prints, arXiv:2406.11502

\bibitem[{{Gardini} {et~al.}(2004){Gardini}, {Rasia}, {Mazzotta}, {Tormen}, {De
  Grandi}, \& {Moscardini}}]{Gardini2004}
{Gardini}, A., {Rasia}, E., {Mazzotta}, P., {et~al.} 2004, \mnras, 351, 505

\bibitem[{Henderson {et~al.}(2016)Henderson, Allison, Austermann, Baildon,
  Battaglia, Beall, Becker, De~Bernardis, Bond, Calabrese,
  {et~al.}}]{henderson2016advanced}
Henderson, S., Allison, R., Austermann, J., {et~al.} 2016, Journal of Low
  Temperature Physics, 184, 772

\bibitem[{{Henning} {et~al.}(2018){Henning}, {Sayre}, {Reichardt}, {Ade},
  {Anderson}, {Austermann}, {Beall}, {Bender}, {Benson}, {Bleem}, {Carlstrom},
  {Chang}, {Chiang}, {Cho}, {Citron}, {Corbett Moran}, {Crawford}, {Crites},
  {de Haan}, {Dobbs}, {Everett}, {Gallicchio}, {George}, {Gilbert},
  {Halverson}, {Harrington}, {Hilton}, {Holder}, {Holzapfel}, {Hoover}, {Hou},
  {Hrubes}, {Huang}, {Hubmayr}, {Irwin}, {Keisler}, {Knox}, {Lee}, {Leitch},
  {Li}, {Lowitz}, {Manzotti}, {McMahon}, {Meyer}, {Mocanu}, {Montgomery},
  {Nadolski}, {Natoli}, {Nibarger}, {Novosad}, {Padin}, {Pryke}, {Ruhl},
  {Saliwanchik}, {Schaffer}, {Sievers}, {Smecher}, {Stark}, {Story}, {Tucker},
  {Vanderlinde}, {Veach}, {Vieira}, {Wang}, {Whitehorn}, {Wu}, \&
  {Yefremenko}}]{Henning2018}
{Henning}, J.~W., {Sayre}, J.~T., {Reichardt}, C.~L., {et~al.} 2018, \apj, 852,
  97

\bibitem[{{Hersbach} {et~al.}(2020){Hersbach}, {Bell}, {Berrisford},
  {Hirahara}, {Hor\'anyi}, J., {Nicolas}, {Peubey}, {Radu}, {Schepers},
  {Simmons}, {Soci}, {Abdalla}, {Abellan}, {Balsamo}, {Bechtold}, G., J.,
  {Bonavita}, {De Chiara}, {Dahlgren}, {Dee}, {Diamantakis}, {Dragani},
  {Flemming}, {Forbes}, {Fuentes}, A., {Haimberger}, {Healy}, {Hogan},
  {H\'olm}, {Janiskov\'a}, {Keeley}, {Laloyaux}, {Lopez}, {Lupu}, {Radnoti},
  {de Rosnay}, {Rozum}, {Vamborg}, {Villaume}, \& {Thépaut}}]{ERA5}
{Hersbach}, H., {Bell}, B., {Berrisford}, P., {et~al.} 2020, QJRMS, 146:730,
  1999

\bibitem[{{Hill} {et~al.}(2018){Hill}, {Bruno}, {Simon}, {Ali}, {Arnold},
  {Ashton}, {Barron}, {Bryan}, {Chinone}, {Coppi}, {Crowley}, {Cukierman},
  {Dicker}, {Dunkley}, {Fabbian}, {Galitzki}, {Gallardo}, {Gudmundsson},
  {Hubmayr}, {Keating}, {Kusaka}, {Lee}, {Matsuda}, {Mauskopf}, {McMahon},
  {Niemack}, {Puglisi}, {Sathyanarayana Rao}, {Salatino}, {Sierra}, {Staggs},
  {Suzuki}, {Teply}, {Ullom}, {Westbrook}, {Xu}, \& {Zhu}}]{Hill2018}
{Hill}, C.~A., {Bruno}, S. M.~M., {Simon}, S.~M., {et~al.} 2018, in Society of
  Photo-Optical Instrumentation Engineers (SPIE) Conference Series, Vol. 10708,
  Millimeter, Submillimeter, and Far-Infrared Detectors and Instrumentation for
  Astronomy IX, ed. J.~{Zmuidzinas} \& J.-R. {Gao}, 1070842

\bibitem[{{Huijten} {et~al.}(2022){Huijten}, {Roelvink}, {Brackenhoff},
  {Taniguchi}, {Bakx}, {Marthi}, {Zaalberg}, {Doing}, {Baselmans}, {Chin},
  {Huiting}, {Karatsu}, {Laguna}, {Tamura}, {Takekoshi}, {Yates}, {van Hoven},
  \& {Endo}}]{Huijten2022}
{Huijten}, E., {Roelvink}, Y., {Brackenhoff}, S.~A., {et~al.} 2022, Journal of
  Astronomical Telescopes, Instruments, and Systems, 8, 028005

\bibitem[{Jia {et~al.}(2015)Jia, Cai, Wang, \& Basden}]{jia2015simulation}
Jia, P., Cai, D., Wang, D., \& Basden, A. 2015, Monthly Notices of the Royal
  Astronomical Society, 447, 3467

\bibitem[{{Kaercher} \& {Baars}(2000)}]{Kaercher2000}
{Kaercher}, H.~J. \& {Baars}, J.~W. 2000, in Society of Photo-Optical
  Instrumentation Engineers (SPIE) Conference Series, Vol. 4015, Radio
  Telescopes, ed. H.~R. {Butcher}, 155--168

\bibitem[{{Kiselev} {et~al.}(2024){Kiselev}, {Reichert}, \&
  {Mroczkowski}}]{Kiselev2024}
{Kiselev}, A., {Reichert}, M., \& {Mroczkowski}, T. 2024, arXiv e-prints,
  arXiv:2404.17311

\bibitem[{{Kisner} {et~al.}(2023){Kisner}, {Keskitalo}, {Zonca}, {Madsen},
  {Puglisi}, {Demeure}, \& {Cheung}}]{Kisner2023}
{Kisner}, T., {Keskitalo}, R., {Zonca}, A., {et~al.} 2023, {TOAST: Time Ordered
  Astrophysics Scalable Tools}, Astrophysics Source Code Library, record
  ascl:2307.022

\bibitem[{{Klaassen} {et~al.}(2020){Klaassen}, {Mroczkowski}, {Cicone},
  {Hatziminaoglou}, {Sartori}, {De Breuck}, {Bryan}, {Dicker}, {Duran},
  {Groppi}, {Kaercher}, {Kawabe}, {Kohno}, \& {Geach}}]{Klaassen2020}
{Klaassen}, P.~D., {Mroczkowski}, T.~K., {Cicone}, C., {et~al.} 2020, in
  Society of Photo-Optical Instrumentation Engineers (SPIE) Conference Series,
  Vol. 11445, Ground-based and Airborne Telescopes VIII, ed. H.~K. {Marshall},
  J.~{Spyromilio}, \& T.~{Usuda}, 114452F

\bibitem[{{Lay}(1997)}]{lay:1997}
{Lay}, O.~P. 1997, \aaps, 122

\bibitem[{{Lee} {et~al.}(2024){Lee}, {Schimek}, {Cicone}, {Andreani},
  {Popping}, {Sommovigo}, {Appleton}, {Bischetti}, {Cantalupo}, {Chen},
  {Dannerbauer}, {De Breuck}, {Di Mascolo}, {Emonts}, {Hatziminaoglou},
  {Pensabene}, {Rizzo}, {Rybak}, {Shen}, {Lundgren}, {Booth}, {Klaassen},
  {Mroczkowski}, {Cordiner}, {Johnstone}, {van Kampen}, {Liu}, {Maccarone},
  {Saintonge}, {Smith}, {Thelen}, \& {Wedemeyer}}]{Lee2024}
{Lee}, M., {Schimek}, A., {Cicone}, C., {et~al.} 2024, Open Research Europe, 4,
  117

\bibitem[{{MacCrann} {et~al.}(2024){MacCrann}, {Sherwin}, {Qu}, {Namikawa},
  {Madhavacheril}, {Abril-Cabezas}, {An}, {Austermann}, {Battaglia},
  {Battistelli}, {Beall}, {Bolliet}, {Bond}, {Cai}, {Calabrese}, {Coulton},
  {Darwish}, {Duff}, {Duivenvoorden}, {Dunkley}, {Farren}, {Ferraro}, {Golec},
  {Guan}, {Han}, {Herv{\'\i}as-Caimapo}, {Hill}, {Hilton}, {Hlo{\v{z}}ek},
  {Hubmayr}, {Kim}, {Li}, {Kosowsky}, {Louis}, {McMahon}, {Marques}, {Moodley},
  {Naess}, {Niemack}, {Page}, {Partridge}, {Schaan}, {Sehgal}, {Sif{\'o}n},
  {Wollack}, {Salatino}, {Ullom}, {Van Lanen}, {Van Engelen}, \&
  {Wenzl}}]{MacCrann2024}
{MacCrann}, N., {Sherwin}, B.~D., {Qu}, F.~J., {et~al.} 2024, \apj, 966, 138

\bibitem[{{Madhavacheril} {et~al.}(2024){Madhavacheril}, {Qu}, {Sherwin},
  {MacCrann}, {Li}, {Abril-Cabezas}, {Ade}, {Aiola}, {Alford}, {Amiri},
  {Amodeo}, {An}, {Atkins}, {Austermann}, {Battaglia}, {Battistelli}, {Beall},
  {Bean}, {Beringue}, {Bhandarkar}, {Biermann}, {Bolliet}, {Bond}, {Cai},
  {Calabrese}, {Calafut}, {Capalbo}, {Carrero}, {Challinor}, {Chesmore}, {Cho},
  {Choi}, {Clark}, {C{\'o}rdova Rosado}, {Cothard}, {Coughlin}, {Coulton},
  {Crowley}, {Dalal}, {Darwish}, {Devlin}, {Dicker}, {Doze}, {Duell}, {Duff},
  {Duivenvoorden}, {Dunkley}, {D{\"u}nner}, {Fanfani}, {Fankhanel}, {Farren},
  {Ferraro}, {Freundt}, {Fuzia}, {Gallardo}, {Garrido}, {Givans}, {Gluscevic},
  {Golec}, {Guan}, {Hall}, {Halpern}, {Han}, {Harrison}, {Hasselfield},
  {Healy}, {Henderson}, {Hensley}, {Herv{\'\i}as-Caimapo}, {Hill}, {Hilton},
  {Hilton}, {Hincks}, {Hlo{\v{z}}ek}, {Ho}, {Huber}, {Hubmayr}, {Huffenberger},
  {Hughes}, {Irwin}, {Isopi}, {Jense}, {Keller}, {Kim}, {Knowles}, {Koopman},
  {Kosowsky}, {Kramer}, {Kusiak}, {La Posta}, {Lague}, {Lakey}, {Lee}, {Li},
  {Limon}, {Lokken}, {Louis}, {Lungu}, {MacInnis}, {Maldonado}, {Maldonado},
  {Mallaby-Kay}, {Marques}, {McMahon}, {Mehta}, {Menanteau}, {Moodley},
  {Morris}, {Mroczkowski}, {Naess}, {Namikawa}, {Nati}, {Newburgh}, {Nicola},
  {Niemack}, {Nolta}, {Orlowski-Scherer}, {Page}, {Pandey}, {Partridge},
  {Prince}, {Puddu}, {Radiconi}, {Robertson}, {Rojas}, {Sakuma}, {Salatino},
  {Schaan}, {Schmitt}, {Sehgal}, {Shaikh}, {Sierra}, {Sievers}, {Sif{\'o}n},
  {Simon}, {Sonka}, {Spergel}, {Staggs}, {Storer}, {Switzer}, {Tampier},
  {Thornton}, {Trac}, {Treu}, {Tucker}, {Ullom}, {Vale}, {Van Engelen}, {Van
  Lanen}, {van Marrewijk}, {Vargas}, {Vavagiakis}, {Wagoner}, {Wang}, {Wenzl},
  {Wollack}, {Xu}, {Zago}, \& {Zheng}}]{Madhavacheril2024}
{Madhavacheril}, M.~S., {Qu}, F.~J., {Sherwin}, B.~D., {et~al.} 2024, \apj,
  962, 113

\bibitem[{{Maud} {et~al.}(2023){Maud}, {Asaki}, {Nagai}, {Tsukui}, {Hirota},
  {Fomalont}, {Dent}, {Takahashi}, \& {Phillips}}]{Maud2023}
{Maud}, L.~T., {Asaki}, Y., {Nagai}, H., {et~al.} 2023, \apjs, 267, 24

\bibitem[{{Morris} {et~al.}(2024){Morris}, {Battistelli}, {Bustos}, {Choi},
  {Duivenvoorden}, {Dunkley}, {D{\"u}nner}, {Halpern}, {Guan}, {van Marrewijk},
  {Mroczkowski}, {Naess}, {Niemack}, {Page}, {Partridge}, {Puddu}, {Salatino},
  {Sif{\'o}n}, {Wang}, \& {Wollack}}]{Morris2024}
{Morris}, T.~W., {Battistelli}, E., {Bustos}, R., {et~al.} 2024, arXiv
  e-prints, arXiv:2410.13064

\bibitem[{{Morris} {et~al.}(2022){Morris}, {Bustos}, {Calabrese}, {Choi},
  {Duivenvoorden}, {Dunkley}, {D{\"u}nner}, {Gallardo}, {Hasselfield},
  {Hincks}, {Mroczkowski}, {Naess}, {Niemack}, {Page}, {Partridge}, {Salatino},
  {Staggs}, {Treu}, {Wollack}, \& {Xu}}]{morris2022}
{Morris}, T.~W., {Bustos}, R., {Calabrese}, E., {et~al.} 2022, \prd, 105,
  042004

\bibitem[{{Mroczkowski} {et~al.}(2023){Mroczkowski}, {Cicone}, {Reichert},
  {Gallardo}, {Kaercher}, {Hills}, {Bok}, {Dahl}, {Dubois-Dit-Bonclaude},
  {Kiselev}, {Timpe}, {Zimmerer}, {Dicker}, {Macintosh}, {Klaassen}, \&
  {Niemack}}]{mroczkowski2023}
{Mroczkowski}, T., {Cicone}, C., {Reichert}, M., {et~al.} 2023, in 2023 XXXVth
  General Assembly and Scientific Symposium of the International Union of Radio
  Science (URSI GASS), 174

\bibitem[{Mroczkowski {et~al.}(2024)Mroczkowski, Gallardo, Timpe, Kiselev,
  Groh, Kaercher, Reichert, Cicone, Puddu, dit Bonclaude, Bok, Dahl, Macintosh,
  Dicker, Viole, Sartori, Venegas, Zeyringer, Niemack, Poppi, Olguin,
  Hatziminaoglou, Breuck, Klaassen, Montenegro-Montes, \&
  Zimmerer}]{Mroczkowski2024}
Mroczkowski, T., Gallardo, P.~A., Timpe, M., {et~al.} 2024, arXiv e-prints,
  arXiv:2402.18645

\bibitem[{{Mroczkowski} {et~al.}(2019){Mroczkowski}, {Nagai}, {Basu}, {Chluba},
  {Sayers}, {Adam}, {Churazov}, {Crites}, {Di Mascolo}, {Eckert},
  {Macias-Perez}, {Mayet}, {Perotto}, {Pointecouteau}, {Romero}, {Ruppin},
  {Scannapieco}, \& {ZuHone}}]{Mroczkowski2019}
{Mroczkowski}, T., {Nagai}, D., {Basu}, K., {et~al.} 2019, \ssr, 215, 17

\bibitem[{{Murphy} {et~al.}(2018){Murphy}, {Bolatto}, {Chatterjee}, {Casey},
  {Chomiuk}, {Dale}, {de Pater}, {Dickinson}, {Francesco}, {Hallinan},
  {Isella}, {Kohno}, {Kulkarni}, {Lang}, {Lazio}, {Leroy}, {Loinard},
  {Maccarone}, {Matthews}, {Osten}, {Reid}, {Riechers}, {Sakai}, {Walter}, \&
  {Wilner}}]{Murphy2018}
{Murphy}, E.~J., {Bolatto}, A., {Chatterjee}, S., {et~al.} 2018, in
  Astronomical Society of the Pacific Conference Series, Vol. 517, Science with
  a Next Generation Very Large Array, ed. E.~{Murphy}, 3

\bibitem[{{Naess} \& {Louis}(2023)}]{Naess2023}
{Naess}, S. \& {Louis}, T. 2023, The Open Journal of Astrophysics, 6, 21

\bibitem[{{Paine}(2018)}]{am:2018}
{Paine}, S. 2018, {The am atmospheric model}

\bibitem[{Pardo {et~al.}(2001)Pardo, Cernicharo, \& Serabyn}]{982447}
Pardo, J., Cernicharo, J., \& Serabyn, E. 2001, IEEE Transactions on Antennas
  and Propagation, 49, 1683

\bibitem[{{Plunkett} {et~al.}(2023){Plunkett}, {Hacar}, {Moser-Fischer},
  {Petry}, {Teuben}, {Pingel}, {Kunneriath}, {Takagi}, {Miyamoto}, {Moravec},
  {Suri}, {Hess}, {Hoffman}, \& {Mason}}]{Plunkett2023}
{Plunkett}, A., {Hacar}, A., {Moser-Fischer}, L., {et~al.} 2023, \pasp, 135,
  034501

\bibitem[{{Puddu} {et~al.}(2024){Puddu}, {Gallardo}, {Mroczkowski},
  {Dubois-dit-Bonclaude}, {Groh}, {Kiselev}, {Reichert}, {Timpe}, {Cicone},
  {Kaercher}, \& {D{\"u}nner}}]{Puddu2024}
{Puddu}, R., {Gallardo}, P.~A., {Mroczkowski}, T., {et~al.} 2024, arXiv
  e-prints, arXiv:2406.16602

\bibitem[{{Puglisi} {et~al.}(2021){Puglisi}, {Keskitalo}, {Kisner}, \&
  {Borrill}}]{Puglisi2021}
{Puglisi}, G., {Keskitalo}, R., {Kisner}, T., \& {Borrill}, J.~D. 2021,
  Research Notes of the American Astronomical Society, 5, 137

\bibitem[{{Qu} {et~al.}(2024){Qu}, {Sherwin}, {Madhavacheril}, {Han},
  {Crowley}, {Abril-Cabezas}, {Ade}, {Aiola}, {Alford}, {Amiri}, {Amodeo},
  {An}, {Atkins}, {Austermann}, {Battaglia}, {Battistelli}, {Beall}, {Bean},
  {Beringue}, {Bhandarkar}, {Biermann}, {Bolliet}, {Bond}, {Cai}, {Calabrese},
  {Calafut}, {Capalbo}, {Carrero}, {Carron}, {Challinor}, {Chesmore}, {Cho},
  {Choi}, {Clark}, {C{\'o}rdova Rosado}, {Cothard}, {Coughlin}, {Coulton},
  {Dalal}, {Darwish}, {Devlin}, {Dicker}, {Doze}, {Duell}, {Duff},
  {Duivenvoorden}, {Dunkley}, {D{\"u}nner}, {Fanfani}, {Fankhanel}, {Farren},
  {Ferraro}, {Freundt}, {Fuzia}, {Gallardo}, {Garrido}, {Gluscevic}, {Golec},
  {Guan}, {Halpern}, {Harrison}, {Hasselfield}, {Healy}, {Henderson},
  {Hensley}, {Herv{\'\i}as-Caimapo}, {Hill}, {Hilton}, {Hilton}, {Hincks},
  {Hlo{\v{z}}ek}, {Ho}, {Huber}, {Hubmayr}, {Huffenberger}, {Hughes}, {Irwin},
  {Isopi}, {Jense}, {Keller}, {Kim}, {Knowles}, {Koopman}, {Kosowsky},
  {Kramer}, {Kusiak}, {La Posta}, {Lague}, {Lakey}, {Lee}, {Li}, {Li}, {Limon},
  {Lokken}, {Louis}, {Lungu}, {MacCrann}, {MacInnis}, {Maldonado}, {Maldonado},
  {Mallaby-Kay}, {Marques}, {McMahon}, {Mehta}, {Menanteau}, {Moodley},
  {Morris}, {Mroczkowski}, {Naess}, {Namikawa}, {Nati}, {Newburgh}, {Nicola},
  {Niemack}, {Nolta}, {Orlowski-Scherer}, {Page}, {Pandey}, {Partridge},
  {Prince}, {Puddu}, {Radiconi}, {Robertson}, {Rojas}, {Sakuma}, {Salatino},
  {Schaan}, {Schmitt}, {Sehgal}, {Shaikh}, {Sierra}, {Sievers}, {Sif{\'o}n},
  {Simon}, {Sonka}, {Spergel}, {Staggs}, {Storer}, {Switzer}, {Tampier},
  {Thornton}, {Trac}, {Treu}, {Tucker}, {Ullom}, {Vale}, {Van Engelen}, {Van
  Lanen}, {van Marrewijk}, {Vargas}, {Vavagiakis}, {Wagoner}, {Wang}, {Wenzl},
  {Wollack}, {Xu}, {Zago}, \& {Zheng}}]{Qu2024}
{Qu}, F.~J., {Sherwin}, B.~D., {Madhavacheril}, M.~S., {et~al.} 2024, \apj,
  962, 112

\bibitem[{{Ramasawmy} {et~al.}(2022){Ramasawmy}, {Klaassen}, {Cicone},
  {Mroczkowski}, {Chen}, {Cornish}, {da Cunha}, {Hatziminaoglou}, {Johnstone},
  {Liu}, {Perrott}, {Schimek}, {Stanke}, \& {Wedemeyer}}]{Ramasawmy2022}
{Ramasawmy}, J., {Klaassen}, P.~D., {Cicone}, C., {et~al.} 2022, in Society of
  Photo-Optical Instrumentation Engineers (SPIE) Conference Series, Vol. 12190,
  Millimeter, Submillimeter, and Far-Infrared Detectors and Instrumentation for
  Astronomy XI, ed. J.~{Zmuidzinas} \& J.-R. {Gao}, 1219007

\bibitem[{{Rasia} {et~al.}(2015){Rasia}, {Borgani}, {Murante}, {Planelles},
  {Beck}, {Biffi}, {Ragone-Figueroa}, {Granato}, {Steinborn}, \&
  {Dolag}}]{Rasia015}
{Rasia}, E., {Borgani}, S., {Murante}, G., {et~al.} 2015, \apjl, 813, L17

\bibitem[{{Rasia} {et~al.}(2008){Rasia}, {Mazzotta}, {Bourdin}, {Borgani},
  {Tornatore}, {Ettori}, {Dolag}, \& {Moscardini}}]{Rasia2008}
{Rasia}, E., {Mazzotta}, P., {Bourdin}, H., {et~al.} 2008, \apj, 674, 728

\bibitem[{{Reichert} {et~al.}(2024){Reichert}, {Timpe}, {Kaercher},
  {Mroczkowski}, {Groh}, {Kiselev}, {Cicone}, {Gallardo}, {Puddu}, \&
  {Klaasen}}]{Reichert2024}
{Reichert}, M., {Timpe}, M., {Kaercher}, H., {et~al.} 2024, in Society of
  Photo-Optical Instrumentation Engineers (SPIE) Conference Series, Vol. 13094,
  Ground-based and Airborne Telescopes X, ed. H.~K. {Marshall},
  J.~{Spyromilio}, \& T.~{Usuda}, 130941U

\bibitem[{Rocklin(2015)}]{rocklin2015dask}
Rocklin, M. 2015, Dask: Parallel computation with blocked algorithms and task
  scheduling

\bibitem[{{Romero} {et~al.}(2020){Romero}, {Sievers}, {Ghirardini}, {Dicker},
  {Giacintucci}, {Mroczkowski}, {Mason}, {Sarazin}, {Devlin}, {Gaspari},
  {Battaglia}, {Hilton}, {Bulbul}, {Lowe}, \& {Stanchfield}}]{Romero2020}
{Romero}, C.~E., {Sievers}, J., {Ghirardini}, V., {et~al.} 2020, \apj, 891, 90

\bibitem[{Ruze(1966)}]{Ruze}
Ruze, J. 1966, Proceedings of the IEEE, 54, 633

\bibitem[{{Schimek} {et~al.}(2024){Schimek}, {Decataldo}, {Shen}, {Cicone},
  {Baumschlager}, {van Kampen}, {Klaassen}, {Madau}, {Di Mascolo}, {Mayer},
  {Montoya Arroyave}, {Mroczkowski}, \& {Warraich}}]{Schimek+24}
{Schimek}, A., {Decataldo}, D., {Shen}, S., {et~al.} 2024, \aap, 682, A98

\bibitem[{{Sehgal} {et~al.}(2019){Sehgal}, {Aiola}, {Akrami}, {Basu},
  {Boylan-Kolchin}, {Bryan}, {Clesse}, {Cyr-Racine}, {Di Mascolo}, {Dicker},
  {Essinger-Hileman}, {Ferraro}, {Fuller}, {Han}, {Hasselfield}, {Holder},
  {Jain}, {Johnson}, {Johnson}, {Klaassen}, {Madhavacheril}, {Mauskopf},
  {Meerburg}, {Meyers}, {Mroczkowski}, {M{\"u}nchmeyer}, {Naess}, {Nagai},
  {Namikawa}, {Newburgh}, {Nguyen}, {Niemack}, {Oppenheimer}, {Pierpaoli},
  {Schaan}, {Slosar}, {Spergel}, {Switzer}, {van Engelen}, \&
  {Wollack}}]{CMBHD2019}
{Sehgal}, N., {Aiola}, S., {Akrami}, Y., {et~al.} 2019, in Bulletin of the
  American Astronomical Society, Vol.~51, 6

\bibitem[{{Selig} {et~al.}(2013){Selig}, {Bell}, {Junklewitz}, {Oppermann},
  {Reinecke}, {Greiner}, {Pachajoa}, \& {En{\ss}lin}}]{Selig2014}
{Selig}, M., {Bell}, M.~R., {Junklewitz}, H., {et~al.} 2013, \aap, 554, A26

\bibitem[{{Sunyaev} \& {Zeldovich}(1970)}]{Sunyaev1970}
{Sunyaev}, R.~A. \& {Zeldovich}, Y.~B. 1970, \apss, 7, 3

\bibitem[{{Sunyaev} \& {Zeldovich}(1972)}]{Sunyaev1972}
{Sunyaev}, R.~A. \& {Zeldovich}, Y.~B. 1972, Comments on Astrophysics and Space
  Physics, 4, 173

\bibitem[{Swetz {et~al.}(2011)Swetz, Ade, Amiri, Appel, Battistelli, Burger,
  Chervenak, Devlin, Dicker, Doriese, Dünner, Essinger-Hileman, Fisher,
  Fowler, Halpern, Hasselfield, Hilton, Hincks, Irwin, Jarosik, Kaul, Klein,
  Lau, Limon, Marriage, Marsden, Martocci, Mauskopf, Moseley, Netterfield,
  Niemack, Nolta, Page, Parker, Staggs, Stryzak, Switzer, Thornton, Tucker,
  Wollack, \& Zhao}]{Swetz_2011}
Swetz, D.~S., Ade, P. A.~R., Amiri, M., {et~al.} 2011, The Astrophysical
  Journal Supplement Series, 194, 41

\bibitem[{Tatarski(1961)}]{tatarski}
Tatarski, V.~I. 1961, Wave Propagation in a Turbulent Medium (McGraw-Hill, NY)

\bibitem[{{The CMB-HD Collaboration} {et~al.}(2022){The CMB-HD Collaboration},
  {:}, {Aiola}, {Akrami}, {Basu}, {Boylan-Kolchin}, {Brinckmann}, {Bryan},
  {Casey}, {Chluba}, {Clesse}, {Cyr-Racine}, {Di Mascolo}, {Dicker},
  {Essinger-Hileman}, {Farren}, {Fedderke}, {Ferraro}, {Fuller}, {Galitzki},
  {Gluscevic}, {Grin}, {Han}, {Hasselfield}, {Hlozek}, {Holder}, {Hotinli},
  {Jain}, {Johnson}, {Johnson}, {Klaassen}, {MacInnis}, {Madhavacheril},
  {Mandal}, {Mauskopf}, {Meerburg}, {Meyers}, {Miranda}, {Mroczkowski},
  {Mukherjee}, {Munchmeyer}, {Munoz}, {Naess}, {Nagai}, {Namikawa}, {Newburgh},
  {Nguyen}, {Niemack}, {Oppenheimer}, {Pierpaoli}, {Raghunathan}, {Schaan},
  {Sehgal}, {Sherwin}, {Simon}, {Slosar}, {Smith}, {Spergel}, {Switzer},
  {Trivedi}, {Tsai}, {van Engelen}, {Wandelt}, {Wollack}, \& {Wu}}]{CMB-HD2022}
{The CMB-HD Collaboration}, {:}, {Aiola}, S., {et~al.} 2022, arXiv e-prints,
  arXiv:2203.05728

\bibitem[{Thornton {et~al.}(2016)Thornton, Ade, Aiola, Angil{\`{e}}, Amiri,
  Beall, Becker, Cho, Choi, Corlies, Coughlin, Datta, Devlin, Dicker, Dünner,
  Fowler, Fox, Gallardo, Gao, Grace, Halpern, Hasselfield, Henderson, Hilton,
  Hincks, Ho, Hubmayr, Irwin, Klein, Koopman, Li, Louis, Lungu, Maurin,
  McMahon, Munson, Naess, Nati, Newburgh, Nibarger, Niemack, Niraula, Nolta,
  Page, Pappas, Schillaci, Schmitt, Sehgal, Sievers, Simon, Staggs, Tucker,
  Uehara, van Lanen, Ward, \& Wollack}]{Thornton_2016}
Thornton, R.~J., Ade, P. A.~R., Aiola, S., {et~al.} 2016, The Astrophysical
  Journal Supplement Series, 227, 21

\bibitem[{Tokovinin(2002)}]{tokovinin2002differential}
Tokovinin, A. 2002, Publications of the Astronomical Society of the Pacific,
  114, 1156

\bibitem[{{Wells} {et~al.}(1981){Wells}, {Greisen}, \& {Harten}}]{Wells1981}
{Wells}, D.~C., {Greisen}, E.~W., \& {Harten}, R.~H. 1981, \aaps, 44, 363

\bibitem[{{White} {et~al.}(2022){White}, {Ghigo}, {Prestage}, {Frayer},
  {Maddalena}, {Wallace}, {Brandt}, {Egan}, {Nelson}, \& {Ray}}]{White2022}
{White}, E., {Ghigo}, F.~D., {Prestage}, R.~M., {et~al.} 2022, \aap, 659, A113

\bibitem[{{ZuHone} \& {Hallman}(2016)}]{ZuHone2016}
{ZuHone}, J.~A. \& {Hallman}, E.~J. 2016, {pyXSIM: Synthetic X-ray observations
  generator}, Astrophysics Source Code Library, record ascl:1608.002

\bibitem[{{ZuHone} {et~al.}(2023){ZuHone}, {Vikhlinin}, {Tremblay}, {Randall},
  {Andrade-Santos}, \& {Bourdin}}]{ZuHone2023}
{ZuHone}, J.~A., {Vikhlinin}, A., {Tremblay}, G.~R., {et~al.} 2023, {SOXS:
  Simulated Observations of X-ray Sources}, Astrophysics Source Code Library,
  record ascl:2301.024

\end{thebibliography}

\end{document}